\pdfoutput=1  
\documentclass[11pt]{article}

\usepackage[a4paper,margin=1in]{geometry}
\usepackage{amsmath,amssymb,amsthm}
\usepackage{graphicx}
\usepackage{flafter}
\usepackage{booktabs}
\usepackage{xcolor}
\usepackage[numbers,sort&compress]{natbib}
\usepackage[colorlinks=true,linkcolor=blue,citecolor=blue,urlcolor=blue]{hyperref}
\usepackage{tabularx}
\usepackage{tikz}
\usetikzlibrary{arrows.meta,shapes.geometric,shadows,backgrounds,fit}
\usepackage{fontawesome5}
\usepackage{pgfplots}
\pgfplotsset{compat=1.17}
\usepgfplotslibrary{groupplots}
\definecolor{plotblue}{HTML}{1F77B4}
\definecolor{plotorange}{HTML}{FF7F0E}
\definecolor{plotgreen}{HTML}{2CA02C}
\definecolor{plotred}{HTML}{D62728}
\pgfplotsset{
  report/.style={
    font=\footnotesize,
    line width=0.5pt,
    tick style={line width=0.4pt},
    grid=major, grid style={gray!20},
    legend cell align=left,
    legend style={font=\footnotesize, draw=gray!50, fill=white, fill opacity=0.9,
      text opacity=1, row sep=-1pt},
    every axis plot/.append style={line width=0.9pt},
    xlabel near ticks, ylabel near ticks,
  },
  repeats/.style={only marks, mark=*, mark size=0.8pt, opacity=0.45, forget plot},
}

\newcommand{\code}[1]{\colorbox{gray!12}{\texttt{\vphantom{Ag}#1}}}

\begin{document}
\thispagestyle{plain}

\noindent
\includegraphics[height=1.3cm]{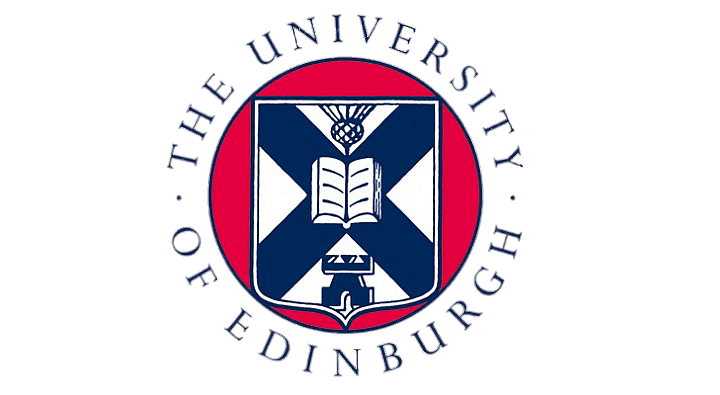}%
\par\vspace{4pt}
\noindent\rule{\textwidth}{0.4pt}

\vspace{3.5em}

\renewcommand{\thefootnote}{\fnsymbol{footnote}}
\begin{center}
  {\LARGE Isabelle Pool Server: a pooled verification service\\for prover training and multi-agent proving\par}
  \vspace{1.6em}
  {\large
    Xuanwei Ren\footnotemark[1]\quad
    Wenda Li\footnotemark[2]\par}
  \vspace{1.6em}
\end{center}
\footnotetext[1]{School of Informatics, University of Edinburgh, UK.
  Email: \texttt{s2415809@ed.ac.uk}}
\footnotetext[2]{School of Informatics, University of Edinburgh, UK.
  Email: \texttt{wenda.li@ed.ac.uk}}
\renewcommand{\thefootnote}{\arabic{footnote}}
\setcounter{footnote}{0}

\vspace{1em}

\begin{abstract}
Isabelle is now driven by machines as well as by people. Training loops verify thousands of candidate proof steps per hour, and autonomous agents reach the prover over the Model Context Protocol, several at a time on one host. Isabelle's own interfaces assume other clients. Its editors each serve one person; its batch build and its server check whole theories for other tools. What both training loops and autonomous agents now need is the prover as a service: calls that return within a known time, a failed attempt that costs nothing to undo, and a verdict that says separately whether anything failed and whether the proof is complete.

To provide this, we introduce Isabelle Pool Server (IPS), a verification service with a pool of warm headless Isabelle sessions behind an HTTP API, with session leasing and reclamation, memory-aware admission control, and a time budget on every call. Its verification interface spans single-step execution, block-level proof checking and whole-theory builds, with checkpoints for cheap branching and backtracking, and a build from scratch as the final arbiter. The service runs as a single container with monitoring and is accessible via a Python client, two MCP servers and a raw HTTP API. In our measurements IPS is level with PIDE MCP, a PIDE frontend for agents, at one worker and ahead of it at eight, at its commit of October 2026 (the tool is now PIDE AI). Once its sessions are up it scales to 32 concurrent workers on one host at 24 times the single-worker rate, and under memory pressure it refuses new sessions rather than losing existing ones. IPS is free software and publicly available at \url{https://github.com/Deep-LCF/Isabelle-Pool-Server}.
\end{abstract}

\section{Introduction}

Proof assistants are now driven by machines as well as by people, in two ways. In the \textbf{training scenario}, a loop that builds or evaluates an LLM prover verifies thousands of candidate proof steps per hour. Isabelle provers since LISA have done this through PISA-style read-eval-print loops (REPLs), one Isabelle process per worker~\cite{jiang2021lisa,jiang2022thor,jiang2023dsp,mikula2024magnushammer}, and two of them name the cost of attempts as the limiting factor~\cite{jiang2023dsp,mikula2024magnushammer}. In the \textbf{multi-agent serving scenario}, many coding agents reach the prover over the Model Context Protocol (MCP)~\cite{anthropic2024mcp} and share one host, on which memory decides how many Isabelle sessions can run. Lean has infrastructure for both: Kimina Lean Server pools REPL workers for reinforcement learning~\cite{santos2025kimina,deepseek2024proverv15}, and AXLE hosts isolated, strictly checked verification for agents and training alike~\cite{xin2026axle}. Isabelle's own interfaces assume other clients. Its editors each serve one person, and \code{isabelle server} checks whole theories for other tools; both are built on PIDE, the document model of the system~\cite{wenzel2014pide}. \code{isabelle build} checks whole sessions in batch. The interfaces that training loops and agents use come from outside the distribution. The loops use the REPLs above. Agents use one of several MCP servers, which differ in how they reach Isabelle: AutoCorrode's Isabelle/Q~\cite{awslabs2026autocorrode} runs inside a person's Isabelle/jEdit session; PIDE MCP~\cite{kappelmann2026pidemcp}, since renamed PIDE AI, drives sessions of its own without an editor; and Isabelle-MCP~\cite{xu2026isabellemcp} goes through the language server. Each meets some of the needs below. None, as far as we know, manages the resources of the host that its clients share.

The two scenarios need the same things of a prover service. A call must return within a known time: a training loop bounds every step so that no attempt stalls a worker, and an agent's harness abandons a tool call that outlasts its timeout. A failed attempt must cost nothing to undo: a loop that branches at a proof state discards most of what it tries, and an agent whose step fails has to get back to where it was. And the answer to whether a proof is complete must be reliable, because a verifier that accepts an unfinished proof corrupts a reward~\cite{deepseek2025proverv2,xin2026axle} and tells an agent that it has finished when it has not. Where many clients share one host, two more things are needed: each client's work kept apart from the others', and a component that decides who gets a session, when it is taken back, and what happens when memory runs out.

IPS is a service of this kind. Like PIDE MCP, it is a direct PIDE frontend: it drives headless sessions through Isabelle/Scala, with no editor between. What it adds is a resource-management layer, which makes those decisions. It keeps a pool of warm sessions, \emph{leases} them to clients over HTTP, reclaims idle and abandoned leases, and gates admission on memory, with a wall-clock budget on every wait. Because a PIDE session is a document and not a command stream~\cite{wenzel2012repl}, IPS's tools can roll back a failed attempt, restore a \emph{checkpoint} exactly and check a \emph{chunk}'s proofs in parallel; a batch build gives the final verdict. Training harnesses use a Python client; agents use one of two bundled MCP servers, one that verifies submitted proof chunks and one that follows a theory file the agent edits on disk.

IPS is built for these two scenarios. We state what each needs from the service as metrics before reporting on them (Section~\ref{sec:readers}). Our contributions are:
\begin{enumerate}
  \item \textbf{A resource-management layer on a direct PIDE frontend.}
  A pool of leased sessions with memory-gated admission, and on it the tools of a verification service: budgets, rollback of a failed chunk as a unit, checkpoints, a batch build as arbiter, a Python client and two MCP servers (Section~\ref{sec:arch}).
  \item \textbf{Measurements for both scenarios.} For the training scenario: throughput, rollout rate and branch cost on the HOL-Analysis library, with two checks of the implementation, the verdict flags and the bounded step. For the multi-agent serving scenario: memory per session, acquisition latency, behaviour under a memory limit and crash containment. All were taken on one 128\,GB host, and PIDE MCP, at commit \code{dd46315}, was measured under the same conditions for memory, throughput, latency and crash containment (Sections~\ref{sec:training}, \ref{sec:serving} and~\ref{sec:comparison}).
  \item \textbf{A comparison with PIDE MCP at one commit.} Where the two direct PIDE frontends differed at the commit we measured, which of the differences are tooling that would transfer, and a feature inventory (Section~\ref{sec:comparison}; the full table is in the repository).
  \item \textbf{An open implementation and a benchmark.} The service\footnote{\url{https://github.com/Deep-LCF/Isabelle-Pool-Server}} and the harness that drove both tools through one procedure are public, and so are the raw results, in a results repository\footnote{\url{https://github.com/FullBackward/Isabelle-Pool-Server-results}} (Appendix~\ref{app:bench}).
\end{enumerate}
Three things here are useful to readers who will never run IPS. The sizing numbers of Section~\ref{sec:serving} apply to any shared Isabelle2026 host. The memory cost of treating a whole theory as visible to PIDE, found in PIDE MCP at the measured commit and reproduced on IPS (Section~\ref{sec:comparison}), is a finding for any PIDE frontend. And the harness is a benchmark a third connector can plug into; its verdict-flag result, an interactive PIDE verdict matching \code{isabelle build} on all 890 checks, is evidence for any frontend that reads its verdict from PIDE, not for IPS alone. Section~\ref{sec:position} first places IPS among Isabelle connectors; Section~\ref{sec:limitations} lists the limits.

\section{Isabelle connectors and where IPS sits}
\label{sec:position}

A connector between a program and Isabelle answers three questions, which this section keeps apart:
\begin{enumerate}
    \item \emph{How the client reaches the server}: the server's interface (REPL, LSP, command line, MCP, HTTP).

    \item \emph{How the server reaches Isabelle}: through Isabelle/ML alone, through a protocol such as that of \code{isabelle server} or of the language server, or through the Isabelle/Scala API of PIDE.

    \item \emph{What the server offers}: its tools (state queries, edits, several sessions, reuse of sessions) and its background services (a memory guard, a file watcher).
\end{enumerate}
The second answer fixes an architecture. The third does not: a tool set changes from one commit to the next, and a tool that one server lacks can often be written for it. So we name the version we read of each connector whose tools we describe. For PIDE MCP it is commit \code{dd46315} of 19 October 2026 (Section~\ref{sec:pide-version}).

A \emph{PIDE frontend} is a program that manages one or several PIDE sessions through PIDE's Isabelle/Scala API~\cite{kappelmann2026pidemcp}: Isabelle/jEdit, \code{isabelle server}, PIDE MCP, IPS. A \emph{direct} frontend uses PIDE itself, with no editor between; an \emph{indirect} one goes through an editor's session. Kappelmann's own table calls the two ``headless PIDE'' and ``editor-attached PIDE''. We use the other pair, because ``headless'' is easily read as implying one process, or one session, per agent, which it does not. The pair was suggested to us by Kevin Kappelmann (personal communication, November 2026). Table~\ref{tab:terms} collects the terms the report uses throughout and says whose definition each is: Wenzel's, from the Isabelle manuals~\cite{wenzel2026system,wenzel2026jedit,wenzel2026isar}, Kappelmann's, IPS's for its API, or this report's for its experiments. What the training and multi-agent serving scenarios need of a service is set out as metrics in Table~\ref{tab:metrics}.

\begin{table}[!htb]
\centering\small
\begin{tabularx}{\textwidth}{@{}l >{\raggedright\arraybackslash}X >{\raggedright\arraybackslash}p{30mm}@{}}
\toprule
Term & Meaning & Source \\
\midrule
heap & the persistent heap image of a logic session, which captures the state of its theories; built once by \code{isabelle build} and loaded by every session that starts on it & Wenzel~\cite{wenzel2026system}, \S2 \\
session & an Isabelle/PIDE session with its underlying Isabelle/ML process, started on a heap, as \code{isabelle server}'s \code{session\_start} does & Wenzel~\cite{wenzel2026system}, \S4.4 \\
settled & every command of the node has finished or failed, the node status under which \code{use\_theories} returns & Wenzel~\cite{wenzel2026system}, \S4.3 \\
batch build & \code{isabelle build} of the theory on its heap: the whole-file check, and the final verdict & Wenzel~\cite{wenzel2026system}, \S2 \\
perspective, overlay & the visible range of a node, which decides where PIDE runs its print functions; a query attached to a command without an edit & Wenzel~\cite{wenzel2026jedit}, \S3.1 \\
node & a unit of the PIDE document: a theory, or an auxiliary file such as an ML file; the preprint's ``document'' & Wenzel~\cite{wenzel2014pide} \\
print function & a function PIDE runs by itself on a checked command that is visible or carries an overlay; Isabelle's automatic tools are print functions & Wenzel~\cite{wenzel2014pide} \\
PIDE frontend & a program that manages one or several PIDE sessions through PIDE's Isabelle/Scala API & Kappelmann~\cite{kappelmann2026pidemcp}, \S2 \\
chunk & one document edit of one or more consecutive Isar commands, submitted as a unit & this work \\
checkpoint, restore & a recorded version of the node's document, and the edit that takes its text back to that version & this work \\
lease & one client's exclusive hold on a session; it expires, so an abandoned session is reclaimed & this work \\
census & one replay of the corpus at one worker; its times are the reference for every slowdown & this work \\
topology A, B & PIDE MCP at commit \code{dd46315} as one process holding several sessions for one client; one such process per session & this work \\
\bottomrule
\end{tabularx}
\caption{Terms used throughout the report, and whose definition each is.}
\label{tab:terms}
\end{table}

The connectors in use today are of six kinds. The seventh item below is not a kind of connector but the layer that IPS adds.

\begin{itemize}
    \item \textbf{REPLs on Isabelle/ML.} Examples are PISA and its descendants~\cite{jiang2021lisa} and qIsabelle~\cite{wrochna2024qisabelle}. On the Lean side the Lean REPL~\cite{leanrepl2023} is of the same kind, and Pantograph~\cite{aniva2025pantograph} is a variant built for proof search. Each runs one prover process per worker, takes commands one at a time, and holds the state in memory. They were built for the training scenario and are still its most common harness. Two newer ones serve agents: Isa-Mini's REPL server, on which AoA~\cite{xu2026aoa} runs, and AutoCorrode's Isabelle/REPL (I/R)~\cite{awslabs2026autocorrode}, which can fork a REPL at a point in another. A REPL is sequential by construction, so parallel proof checking is lost~\cite{wenzel2012repl}.

    \item \textbf{Batch build.} The agent edits files and runs \code{isabelle build}, as in the Munkres autoformalisation~\cite{bryant2026munkres}. A build is not incremental: every attempt re-checks the whole development. AutoCorrode's Isabelle/Check (I/C) is a file check that avoids this: it runs on I/R and reprocesses only what a change affects.

    \item \textbf{Isabelle's own server.} \code{isabelle server}, in the distribution since Isabelle2018, is a resident process that manages several logic sessions over a TCP line protocol. It is a direct PIDE frontend, multi-session and bounded by a time limit and a watchdog, but \code{use\_theories} is its one checking command: it works on whole theory files and answers when they are done. There is no command for a single step, for the goal at a point in a proof, or for Sledgehammer.

    \item \textbf{Language server (LSP).} The connector reaches Isabelle through its language server, \code{vscode\_server}, an approach Kappelmann assesses as only partly complete~\cite{kappelmann2026pidemcp}. The stock server lacks PIDE requests that Isabelle-MCP~\cite{xu2026isabellemcp} needs, among them the output at a position and the cancellation of a running check. Early versions of Isabelle-MCP patched the Isabelle distribution to add them; since version 0.3.0 it carries a fork of the server as a component of its own, tied to one Isabelle release. At version 0.7.0 it serves one agent per process. On the Lean side lean-lsp-mcp~\cite{dressler2026leanlspmcp} is of the same kind.
    
    \item \textbf{Indirect PIDE frontend.} Isabelle/Q (I/Q)~\cite{awslabs2026autocorrode} is an MCP server inside Isabelle/jEdit. The agent works in the human's own PIDE session, which requires a running editor and confines it to one session on one machine.
    
    \item \textbf{Direct PIDE frontends.} Examples are PIDE MCP~\cite{kappelmann2026pidemcp} and IPS (this report); \code{isabelle server} above is a third, and AutoCorrode's IC2, which manages headless sessions from the command line, a fourth. At the commit we measured PIDE MCP served one client; its later versions, renamed PIDE AI, let several agents share one server (Section~\ref{sec:pide-version}).

    \item \textbf{A resource-management layer.} As far as we know, none of the Isabelle connectors above, at the versions we read, decides who gets a session, when it is taken back, or what happens when memory runs out. In Lean, Kimina~\cite{santos2025kimina} and AXLE~\cite{xin2026axle} do: they put a pool of prover processes behind HTTP, with warm workers reused, admission, isolation and a bounded call. A pool is not an architecture of its own; it is a set of tools and background services that a multi-session frontend may offer. IPS offers them on a direct PIDE frontend, and is the one pool in either prover whose workers are PIDE sessions and not REPL processes, so that its tools can use the document model for checkpoints, rollback and parallel checking within a chunk.
\end{itemize}

\paragraph{AutoCorrode's tools.} The AutoCorrode repository~\cite{awslabs2026autocorrode}, a verification framework for imperative programs in Isabelle/HOL, carries four of the connectors above: I/Q, I/R, I/C and IC2, the last ``similar to \code{isabelle server} and \code{isabelle client}''. Two further parts concern a human at the editor: Isabelle/Proxy (I/P) runs the prover on a remote machine while Isabelle/jEdit stays local, and Isabelle Assistant is a language-model assistant inside Isabelle/jEdit. We read the repository on 16 November 2026 and measured none of these tools; I/R's forking is the closest relative of IPS's checkpoints (Section~\ref{sec:training-req}).

Nothing in IPS bypasses PIDE: sessions start, load theories and are edited through the API \code{isabelle server} itself uses.\footnote{One reflective call remains, to the private constructor of \code{Server.Context}, so that the backend can invoke the server's own command handlers; packaging the backend as an Isabelle tool would remove it.}

Table~\ref{tab:families} places the connectors on the first two of the three questions that open this section: how the client reaches the server, and how the server reaches Isabelle. The scenarios differ on the third, what the server offers. Interactive work wants the whole PIDE surface, a project of many theories, and calls that do not block; training wants a narrow contract, a bound on every call, and a pool behind it. Nothing in a direct PIDE frontend rules out offering both; no connector we know of does so today. The limit of a REPL, sequential checking, applies to IPS in part: its one-command mode is a REPL and inherits the sequential cost (Section~\ref{sec:modes}); its chunk call submits a block as one PIDE edit with parallel proofs on, and its whole-file check is a batch build.

\begin{table}[!htb]
\centering\footnotesize
\setlength{\tabcolsep}{4pt}
\begin{tabularx}{\textwidth}{@{}>{\raggedright\arraybackslash}p{3.1cm} >{\raggedright\arraybackslash}p{3.0cm} >{\raggedright\arraybackslash}X >{\raggedright\arraybackslash}X@{}}
\toprule
Connector (version read) & Client reaches it by & It reaches Isabelle by & What it keeps between calls \\
\midrule
PISA, qIsabelle & REPL protocol & Isabelle/ML & one prover process per worker, named states \\
I/R (AutoCorrode) & command line, TCP, MCP wrapper & Isabelle/ML & a set of REPLs that can be forked \\
\code{isabelle build} & command line & a batch process & nothing \\
\code{isabelle server} (Isabelle2026) & TCP line protocol & PIDE, direct & several sessions \\
Isabelle-MCP 0.7.0 & MCP & the language server (a fork of \code{vscode\_server}) & one session, one agent per process \\
I/Q (AutoCorrode) & MCP & PIDE, indirect, through Isabelle/jEdit & the editor's session \\
IC2 (AutoCorrode) & command line & PIDE, direct & persistent headless sessions \\
PIDE MCP \code{dd46315} & MCP on standard input and output & PIDE, direct & several sessions, one client \\
PIDE AI 2026 & MCP or commands, on standard input and output or TCP & PIDE, direct & several sessions, several clients \\
IPS (this report) & HTTP, MCP, Python client & PIDE, direct & a pool of leased sessions, admitted by memory \\
\bottomrule
\end{tabularx}
\caption{Isabelle connectors: how a client reaches each, how each
  reaches Isabelle (the first two questions of this section), and what
  each keeps between calls. Only PIDE MCP at \code{dd46315} and IPS were
  measured.}
\label{tab:families}
\end{table}

\section{Requirements and metrics}
\label{sec:readers}

\subsection{The training scenario}
\label{sec:training-req}

In the \textbf{training scenario} a loop builds or evaluates an LLM prover. Whether it is reinforcement learning, expert iteration or search, the loop has one shape: sample candidate steps, submit them, learn from the verdicts, repeat. IPS is judged by what it does to that loop, which needs four things from the service.

\paragraph{Verdicts at the sampler's rate.} Thor issues up to 300 model queries per theorem~\cite{jiang2022thor}; DeepSeek-Prover-V2 samples 32 candidate proofs per problem at every reinforcement-learning step~\cite{deepseek2025proverv2}. A Lean pool verifies about 20 whole proofs per second on 64 cores, 0.05\,s per proof amortised~\cite{santos2025kimina}; that is the reference point for an Isabelle service.

\paragraph{Cheap branching at a proof state.} Best-first and tree search in Thor~\cite{jiang2022thor}, Magnushammer~\cite{mikula2024magnushammer} and DeepSeek-Prover-V1.5~\cite{deepseek2024proverv15} all branch. In a REPL a branch resumes from a saved state when the loop kept one, as PISA and qIsabelle allow and as AutoCorrode's I/R does by forking a REPL at a point in another~\cite{awslabs2026autocorrode}, and is otherwise a replay from the start or a fresh process.

\paragraph{A bound on every step.} Every one of these systems bounds each prover call, from 2 seconds in Magnushammer~\cite{mikula2024magnushammer} to 300 in DeepSeek-Prover~\cite{deepseek2024proverv15}, because one looping tactic otherwise removes a worker from the pool.

\paragraph{A verdict that means what the reward assumes.} DeepSeek-Prover-V2 traced inflated benchmark results to a Lean bug that failed to report \code{sorry}~\cite{deepseek2025proverv2}; AXLE's strict verifier exists because compilation alone accepts \code{sorry}, unsound axioms and restated theorems~\cite{xin2026axle}.

\subsection{The multi-agent serving scenario}
\label{sec:serving-req}

In the \textbf{multi-agent serving scenario} several agents or harnesses share one Isabelle host, and whoever runs it usually does not know how many will be active or what they will import. IPS is judged by capacity and by conduct under load, and the scenario needs predictability on four counts.

\paragraph{A known number of sessions and a known wait.} Kimina's motivation was that earlier tools paid an initialisation cost on every verification, and reusing warm workers that had already loaded the same imports halved its verification time~\cite{santos2025kimina}. For Isabelle the cost is larger, on disk as well as in time: PISA's instructions for running many servers give each server its own copy of Isabelle and its \emph{heaps}, 35\,GB of disk with the heaps of the Archive of Formal Proofs, to avoid I/O errors~\cite{jiang2021lisa}. Memory, not cores, limits an Isabelle host, and the wait for a session runs from milliseconds, when one is ready, to minutes, when a heap must first be built (Section~\ref{sec:serving-measured}).

\paragraph{No heap build while a client waits.} An Isabelle project heap takes minutes to build (Section~\ref{sec:serving-measured}), so a request whose heap is not yet built waits minutes rather than seconds. IPS keeps each heap once it is built (Section~\ref{sec:acquire}).

\paragraph{Defined behaviour when memory runs out.} Kimina caps the memory of each worker~\cite{santos2025kimina}. An Isabelle host has the same need, with larger processes.

\paragraph{Each client's work kept apart from the others.} AXLE sandboxes every request~\cite{xin2026axle}, and the Isabelle community has proposed container and VM setups for agents~\cite{schaeffeler2026safe}. IPS isolates by lease rather than by process, which is cheaper and weaker; the isolation measurement of Section~\ref{sec:serving-measured} states what it provides.

\subsection{A third scenario, not served}

A third, the \textbf{interactive serving scenario}, is a single agent working on a project, alone or beside a human, with the whole of PIDE within reach. It is where we take PIDE MCP to be strongest, though not the only use its preprint names: that also speaks of deployment on servers and of background autoformalisation~\cite{kappelmann2026pidemcp}. IPS does not primarily serve this scenario; Sections~\ref{sec:position} and~\ref{sec:comparison} make the comparison.

\subsection{Metrics}

Table~\ref{tab:metrics} turns the questions of each scenario into the metrics reported in Sections~\ref{sec:training} and~\ref{sec:serving}, with the procedure for each in Appendix~\ref{app:bench}. Agent success on a benchmark measures the agent and its prompt rather than the service, so it appears only as an anecdote in Appendix~\ref{app:example}.

\begin{table}[t]
\centering\small
\begin{tabularx}{\textwidth}{@{}>{\raggedright\arraybackslash}X l >{\raggedright\arraybackslash}X@{}}
\toprule
Question & Metric & Measured as \\
\midrule
\multicolumn{3}{@{}l}{\emph{Training scenario} (Section~\ref{sec:training})} \\
How many verified chunks per minute? & throughput & \code{verify\_chunk} calls per minute against worker count, with and without session starts \\
How many rollouts per hour? & rollout rate & scripted whole-proof and step-level loops, rollouts and verified steps per hour against worker count \\
Can I branch and backtrack cheaply? & branch cost & restore time against replaying to the same proof state \\
\midrule
\multicolumn{3}{@{}l}{\emph{Multi-agent serving scenario} (Section~\ref{sec:serving})} \\
How many sessions fit in a host? & capacity & memory per session at steady state, by heap; sessions per 100\,GB \\
How long until a client gets a session? & acquisition latency & a ready session, a session start and a heap build, separately \\
What happens when memory runs out? & pressure & admission refusals, evictions, kernel kills, recovery time \\
Are agents isolated from each other? & isolation & crash containment: do the other sessions answer? \\
\midrule
\multicolumn{3}{@{}l}{\emph{Checks} (Section~\ref{sec:checks})} \\
Do the verdict flags catch what they target? & verdict flags & each flag against the failure it exists to catch, with the batch build as the reference, on the corpus and on broken proofs \\
Can one bad step stall a worker? & bounded step & looping steps returned at their budget; recovery; effect on other workers \\
\bottomrule
\end{tabularx}
\caption{Each scenario's questions, the metric that answers each, and how
  it is measured. The last group checks two properties of the
  implementation rather than measuring a scenario.}
\label{tab:metrics}
\end{table}

\section{Service overview}
\label{sec:arch}

\begin{figure}[!htb]
\centering
\resizebox{\textwidth}{!}{
\newcommand{\iconcard}[5]{%
  \node[#1] (#2) at #3 {\hspace{6mm}\begin{tabular}{@{}l@{}}#5\end{tabular}};
  \node[anchor=west, inner sep=0pt, xshift=4pt, text=plotblue, font=\large] at (#2.west) {#4};}
\begin{tikzpicture}[
  x=1cm, y=1cm, font=\footnotesize,
  card/.style={draw=gray!55, fill=white, rounded corners=3pt, inner sep=4pt, font=\scriptsize,
    drop shadow={shadow xshift=0.5pt, shadow yshift=-0.7pt, opacity=0.16, fill=black}},
  client/.style={card, minimum width=34mm, minimum height=10mm},
  part/.style={card, minimum width=38mm, minimum height=9.5mm},
  proc/.style={card, align=center, minimum width=16mm, minimum height=9mm, fill=green!5},
  faded/.style={proc, draw=gray!50, text=gray!60, dashed, fill=white, drop shadow={opacity=0}},
  panel/.style={fill=#1, rounded corners=6pt, draw=none, inner sep=5pt},
  ptitle/.style={font=\footnotesize\bfseries, text=black!70, anchor=west, inner sep=2pt},
  arr/.style={-{Latex[length=2.2mm]}, line width=0.8pt, draw=black!55},
  lbl/.style={font=\scriptsize, inner sep=1.5pt, text=black!70}
]
\node[ptitle] (ct) at (0.3, 3.55) {Clients};
\iconcard{client}{c1}{(1.9, 2.7)}{\faCogs}{Training harness\\(Python client)}
\iconcard{client}{c2}{(1.9, 1.45)}{\faRobot}{Agent\\(stepwise MCP)}
\iconcard{client}{c3}{(1.9, 0.2)}{\faTerminal}{Agent or tool\\(file MCP, HTTP)}
\node[ptitle] (st) at (4.4, 3.55) {IPS server (FastAPI)};
\iconcard{part}{sm}{(6.45, 2.2)}{\faLayerGroup}{session manager\\pool, leases, LRU}
\iconcard{part}{mg}{(10.65, 2.2)}{\faMemory}{memory gate\\cgroup use, a margin}
\iconcard{part}{hp}{(6.45, 0.7)}{\faDatabase}{heap pool\\project heaps}
\iconcard{part}{bv}{(10.65, 0.7)}{\faCheckDouble}{batch verify\\\texttt{isabelle build}}
\node[ptitle] (it) at (13.7, 3.55) {Isabelle};
\iconcard{card, minimum width=54mm, minimum height=16mm}{gw}{(16.5, 2.45)}{\faJava}{\textbf{Gateway JVM} (Py4J, Scala)\\one headless PIDE session\\per pooled session:\\document edits, overlay queries}
\node[proc]  (p1) at (14.7, 0.45) {\textcolor{plotblue}{\faMicrochip}\ prover A\\leased};
\node[proc]  (p2) at (16.5, 0.45) {\textcolor{plotblue}{\faMicrochip}\ prover B\\warm, idle};
\node[faded] (p3) at (18.3, 0.45) {\faMicrochip\ prover C\\starting};
\node[cylinder, shape border rotate=90, aspect=0.12, draw=gray!55, fill=yellow!14,
  minimum width=52mm, minimum height=12mm, font=\scriptsize, align=center,
  drop shadow={shadow xshift=0.5pt, shadow yshift=-0.7pt, opacity=0.16, fill=black}]
  (hd) at (16.5, -1.75) {\textcolor{plotblue}{\faHdd}\ heaps on disk\\HOL, HOL-Analysis, project heaps};
\coordinate (sb) at (sm.west |- c3.south);
\begin{scope}[on background layer]
  \node[panel=gray!8]  (pc) [fit=(ct)(c1)(c3)] {};
  \node[panel=blue!5]  (ps) [fit=(st)(sm)(mg)(hp)(bv)(sb)] {};
  \node[panel=green!5] (pi) [fit=(it)(gw)(p1)(p3)(hd)] {};
\end{scope}
\draw[arr] (c1.east) -- (ps.west |- c1.east);
\draw[arr] (c2.east) -- (ps.west |- c2.east);
\draw[arr] (c3.east) -- (ps.west |- c3.east);
\draw[arr] (ps.east |- gw.west) -- (gw.west) node[midway, above, lbl] {Py4J};
\draw[arr] (gw.south -| p1.north) -- (p1.north);
\draw[arr, dashed] (gw.south -| p2.north) -- (p2.north);
\draw[arr, dashed, draw=gray!60] (gw.south -| p3.north) -- (p3.north);
\draw[arr, dashed] (hd.north -| p3.south) -- (p3.south) node[midway, left, lbl] {loaded from};
\end{tikzpicture}}
\caption{IPS as a system. A training harness or an agent reaches the server through the Python client, one of the two bundled MCP servers or plain HTTP. The server leases sessions, admits new ones by memory, keeps project heaps and runs batch builds; a gateway JVM drives one headless PIDE session for each session of the pool, each with an Isabelle/ML process started from a heap. Arrows show who calls whom; every call returns its result along the same path. The dashed arrow from the heaps shows where a prover is loaded from.}
\label{fig:arch}
\end{figure}
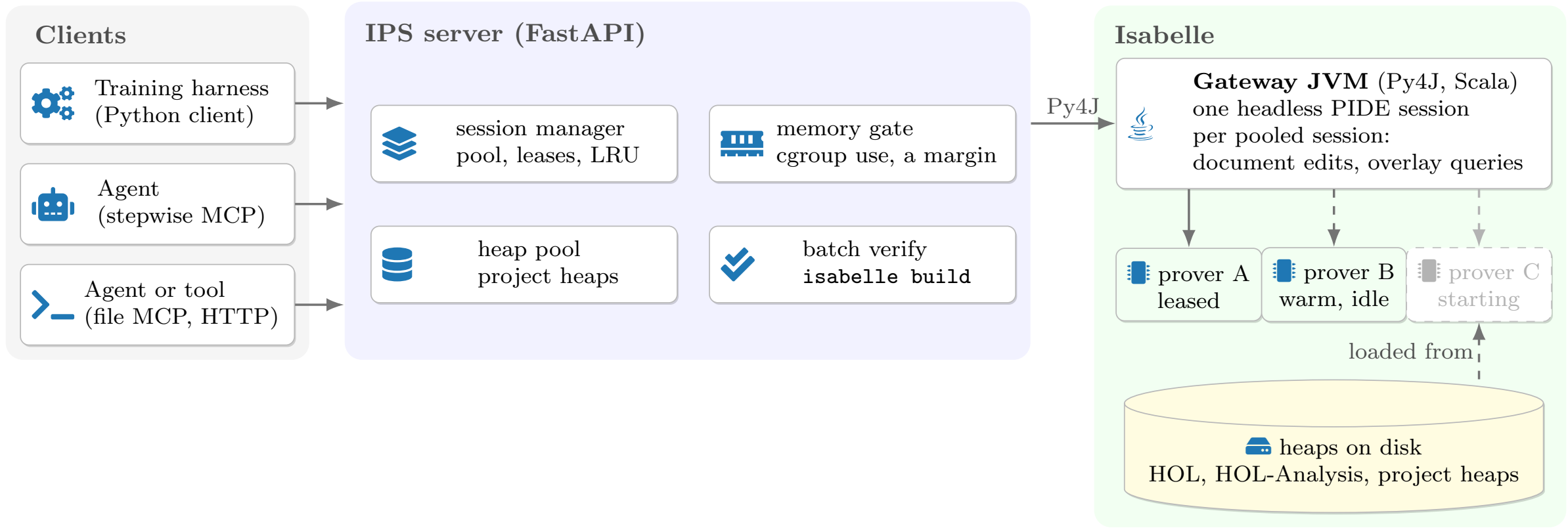

\begin{figure}[!htb]
\centering
\resizebox{\textwidth}{!}{
\begin{tikzpicture}[
  x=1cm, y=1cm,
  box/.style={draw, rounded corners=2pt, align=center, inner sep=3pt, font=\scriptsize},
  dec/.style={box, fill=orange!10, minimum width=24mm, minimum height=9mm},
  outc/.style={box, fill=green!10, minimum width=24mm, minimum height=7.5mm},
  refuse/.style={box, fill=red!7, minimum width=24mm, minimum height=7.5mm},
  arr/.style={-{Latex[length=2.2mm]}, thick},
  lbl/.style={font=\scriptsize, inner sep=1.5pt}
]
\node[dec] (d1) at (1.2, 0)  {idle session for\\these imports?};
\node[dec] (d2) at (5.0, 0)  {heap covering\\the imports?};
\node[dec] (d3) at (8.8, 0)  {memory for\\one more?};
\node[dec] (d4) at (12.6, 0) {pool slot free,\\or one to evict?};
\node[outc] (st) at (16.4, 0) {start a session,\\lease it};
\node[outc]   (le) at (1.2, -1.7)  {lease it};
\node[refuse] (r2) at (5.0, -1.7)  {409: build the\\heap first};
\node[refuse] (r3) at (8.8, -1.7)  {503, \texttt{Retry-After}};
\node[refuse] (r4) at (12.6, -1.7) {503, pool full};
\draw[arr] (d1) -- (d2) node[midway, above, lbl] {no};
\draw[arr] (d2) -- (d3) node[midway, above, lbl] {yes};
\draw[arr] (d3) -- (d4) node[midway, above, lbl] {yes};
\draw[arr] (d4) -- (st) node[midway, above, lbl] {yes};
\draw[arr] (d1) -- (le) node[midway, right, lbl] {yes};
\draw[arr] (d2) -- (r2) node[midway, right, lbl] {no};
\draw[arr] (d3) -- (r3) node[midway, right, lbl] {no};
\draw[arr] (d4) -- (r4) node[midway, right, lbl] {no};
\end{tikzpicture}}
\caption{How a request for a session is resolved, after Kimina's dispatch figure~\cite{santos2025kimina}. A request that finds no warm session needs a heap that covers the imports, memory for one more session once idle ones are reclaimed, and a free slot or an idle session to evict; the first that is missing is the reason it is refused.}
\label{fig:acquire}
\end{figure}
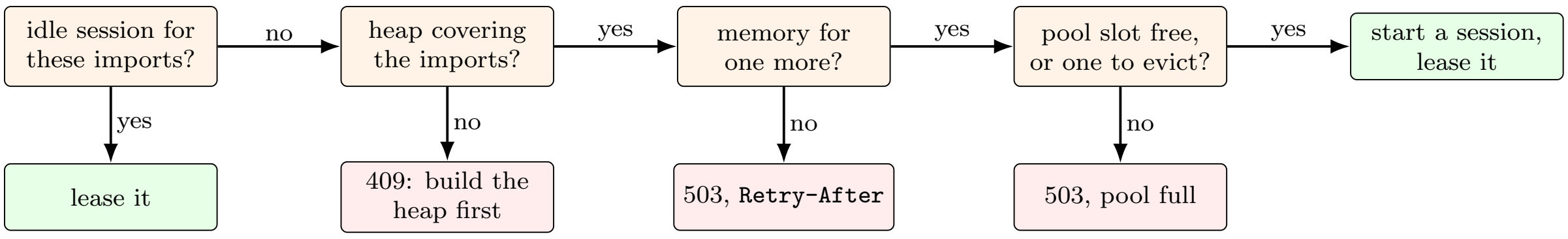

This section describes IPS as a client sees it. Figure~\ref{fig:arch} shows the parts; Figure~\ref{fig:acquire} shows how a request for a session is resolved. A reader who knows PIDE will find the same service described in PIDE's terms in Appendix~\ref{sec:pide-terms}.

\subsection{Sessions and leases}
\label{sec:sessions}

A client works with a \emph{session}: a running Isabelle, loaded with the imports the client asked for, holding one theory the client builds up by submitting text. The client never starts Isabelle and never sees a file. Three properties of a session are fixed: its imports cannot change, its document is one theory rather than a project, and it belongs to one client for the life of its \emph{lease}. A training harness reaches sessions through the Python client; an agent through the \emph{stepwise} MCP server, which verifies chunks the agent submits, or the \emph{file} MCP server, which follows a theory file the agent edits with its own tools.

\subsection{Acquiring a session}
\label{sec:acquire}

The client names the imports it needs. If an idle session with those imports exists, it is leased at once. Otherwise the pool starts one on a built heap, or, if the host is short of memory, refuses with an explanation. Section~\ref{sec:serving-measured} gives the latencies of the two cases, and Section~\ref{sec:training-measured} the cost when 32 sessions start at once. Imports from the Isabelle distribution are always available. Imports from the client's own project need a heap that IPS builds once on request and reuses until the project's sources change. Leases expire and idle sessions are reclaimed, so a client that fails to release does not hold a session indefinitely.

\subsection{Verification modes}
\label{sec:modes}

Table~\ref{tab:modes} sets out the verification modes: one command at a time; a \emph{chunk} (\code{verify\_chunk}), a block of Isar checked as one unit, which is the call a training loop uses; a whole file, rebuilt from scratch in a fresh Isabelle; and, for an agent, a batch of independent chunks, each on a clean session. A submission that fails or times out is \emph{rolled back}: the server removes its text from the theory, so the session is as it was before the call.

Between submissions the client can read goals, open proofs, facts in scope, or a Sledgehammer suggestion without changing the document; take a \emph{checkpoint} and later \emph{restore} it, so a search branches and backtracks without a new session; or roll back its last submission itself.

\begin{table}[!htb]
\centering\footnotesize
\setlength{\tabcolsep}{4pt}
\begin{tabularx}{\textwidth}{@{}>{\raggedright\arraybackslash}p{2.2cm} >{\raggedright\arraybackslash}X >{\raggedright\arraybackslash}X >{\raggedright\arraybackslash}p{1.6cm} >{\raggedright\arraybackslash}p{1.7cm}@{}}
\toprule
Mode & Use it when & Returns & Rolled back on failure & Parallel checking \\
\midrule
One command & the loop wants the proof state after each step & status and messages of that command & yes & no \\
Chunk (\code{verify\_chunk}) & a candidate step or proof is one block & per-command status; \code{success}, \code{proof\_open}, \code{used\_sorry}, \code{timed\_out} & yes, the whole block & within the block \\
Whole file & a verdict must be final & the build's result; cached by content & stateless & the build's \\
Batch (MCP \code{verify\_batch}) & many independent chunks from one agent & one chunk verdict per item & a clean session per item, closed after & across the pool, client-side \\
\bottomrule
\end{tabularx}
\caption{The verification modes.}
\label{tab:modes}
\end{table}

\subsection{Verdict semantics}
\label{sec:verdict}

IPS separates ``did anything error'' from ``is the theorem proved''. A chunk \emph{succeeds} when no command in it errored. It is \emph{proved} when it succeeded, left no proof open, and used no \code{sorry}, Isar's fake proof that pretends to solve the pending claim~\cite{wenzel2026isar}; the client reads this from three flags in the response. When a verdict matters beyond the current loop, the whole-file check is the arbiter: a fresh build depends on nothing the session holds. Section~\ref{sec:checks} tests the flags against it.

\paragraph{What the verdict certifies.} IPS looks for \code{sorry} and \code{oops} as parsed commands, not as words in the text, so a comment or a string can neither hide one nor fake one. The whole-file check runs \code{isabelle build} with Isabelle's defaults, under which \code{sorry} is an error, so a proof the \emph{batch build} accepts is one Isabelle itself accepts. Neither verdict says that the statement proved is the one asked for, and neither rejects a new axiom. A separate command guard covers the second: unless the operator switches it off, IPS refuses \code{axiomatization}, along with ML and file commands, in any text it is sent (Appendix~\ref{app:deploy}), so a candidate proof cannot postulate what it then uses. A client that checks whole theories of its own should keep the guard on. A strict verdict that also checks the statement and the oracle set is on the roadmap (Section~\ref{sec:limitations}).

\subsection{Time budgets}

Every call that waits on Isabelle carries a budget, chosen by the client and enforced by the server. A submission that exceeds its budget is rolled back, and the call returns a timeout, so the session stays usable and no submission can hold a session or a worker indefinitely. This is a property of IPS's calls, not of its architecture (Section~\ref{sec:comparison}). Appendix~\ref{sec:pide-terms} says how the server keeps these guarantees in PIDE's terms, and Appendix~\ref{app:deploy} how it is configured.

\section{Training workloads}
\label{sec:training}

This section covers the \textbf{training scenario}: how a loop uses IPS, then what we measured against the requirements of Section~\ref{sec:training-req}.

\subsection{Design}

The loop acquires a session per worker with the imports of its problems and keeps it while the imports stay the same. Each candidate is a chunk under a budget, submitted after a checkpoint. A candidate that fails or times out is rolled back by the server. A candidate that checks without error stays in the document, and the loop either goes on from it to the next step or restores the checkpoint to discard it. A candidate that is rolled back or discarded leaves the state as it was, and no session is started to branch. The reward reads the three flags of Section~\ref{sec:verdict}. A final verdict goes to the whole-file check, and a batch of independent candidates to the stepwise MCP's \code{verify\_batch} (Table~\ref{tab:modes}). Appendix~\ref{app:quickstart} shows the loop's calls in Python.

\subsection{Results}
\label{sec:training-measured}

All numbers come from one series of runs on the host of Section~\ref{sec:serving-measured} (32 cores, 128\,GB, Isabelle2026). Memory is in binary units throughout the report: 1\,MB is $2^{20}$ bytes and 1\,GB is $2^{30}$, as the kernel reports them. The workload is the HOL-Analysis library: its 93 theories apart from the collector \code{Analysis.thy}, split into 11\,984 top-level blocks of 166\,438 commands, each block submitted as one chunk. One replay of the library at one worker, the \emph{census}, proved every block; its times are the reference for every slowdown below. Each theory runs on a session acquired with its own imports, as in a curriculum whose problems differ in imports. No model runs: the candidates are scripted, so the rates measure the server alone, with no model's time in them. $W$ is the number of concurrent workers. A \emph{cell} is one run at one setting, in one repeat. Appendix~\ref{app:bench-training} gives the method; the raw results are in the results repository.

\begin{figure}[!htb]
\centering
\begin{tikzpicture}
\begin{axis}[report, width=10cm, height=6.4cm,
  xmode=log, log basis x=2, log ticks with fixed point,
  xtick={1,2,4,8,16,32}, xmin=0.85, xmax=38,
  ymin=0, ymax=4400,
  xlabel={workers $W$}, ylabel={verified chunks per minute},
  legend pos=north west]
\addplot[gray, densely dotted, line width=0.8pt] coordinates {(1,131.5) (2,263.0) (4,526.0) (8,1052.0) (16,2104.0) (32,4208.0)};
\addlegendentry{linear scaling from $W = 1$}
\addplot[plotblue, mark=*, mark size=1.6pt] coordinates {(1,131.5) (2,280.5) (4,473.6) (8,926.5) (16,1676.6) (32,3181.4)};
\addlegendentry{verify only}
\addplot[plotorange, mark=square*, mark size=1.5pt] coordinates {(1,113.3) (2,182.5) (4,297.1) (8,497.3) (16,789.1) (32,898.8)};
\addlegendentry{with session starts}
\addplot[repeats, plotblue] coordinates {(1,131.6) (1,131.5) (1,107.6) (16,1676.6) (16,1718.3) (16,1557.7) (2,280.5) (2,286.5) (2,229.8) (32,3161.9) (32,3214.3) (32,3181.4) (4,468.8) (4,503.4) (4,473.6) (8,901.9) (8,974.6) (8,926.5)};
\addplot[repeats, plotorange] coordinates {(1,113.3) (1,116.4) (1,100.3) (16,789.1) (16,726.2) (16,799.5) (2,151.3) (2,182.5) (2,196.8) (32,898.8) (32,888.7) (32,920.5) (4,239.5) (4,297.1) (4,360.2) (8,360.7) (8,497.3) (8,528.3)};
\end{axis}
\end{tikzpicture}
\caption{Verified chunks per minute against the number of workers. The dotted line is linear scaling from the single-worker rate. The verify-only line is the rate once sessions are up; it falls below the dotted line as calls slow, from 0.41 to 0.52\,s at the median. The line with session starts also counts the start of each theory's session; the gap between the two is the share of a worker's time spent acquiring a session, 10--14\,\% up to 8 workers and 64\,\% at 32. Lines are the medians of three repeats; the small points are the repeats. Eight theories per worker.}
\label{fig:throughput}
\end{figure}
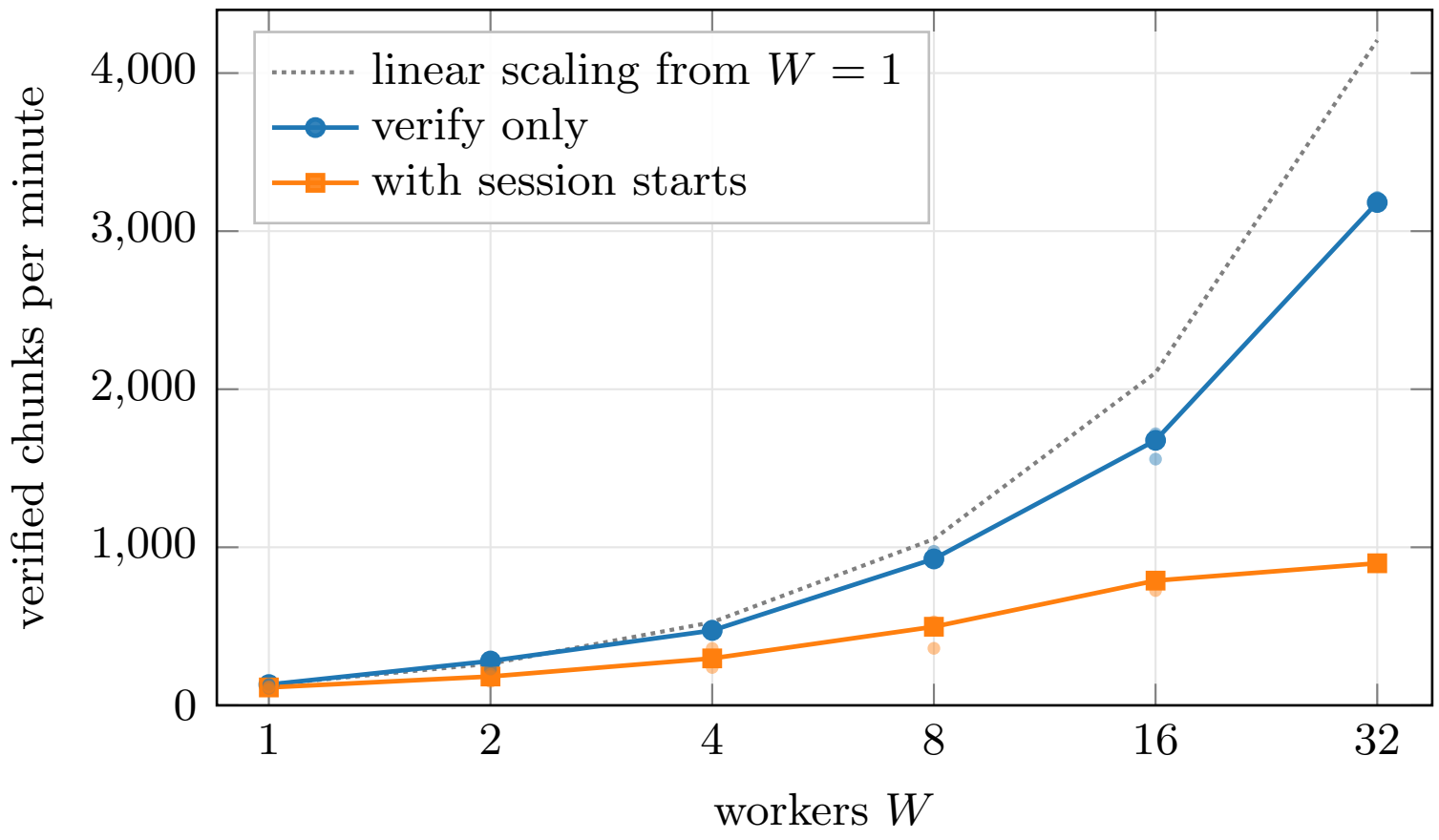

\paragraph{Throughput.} In the throughput runs $W$ workers, from 1 to 32, each replay eight theories block by block with \code{verify\_chunk}, every theory on a session started for it, with three repeats per $W$ (Figure~\ref{fig:throughput}). We give the rate twice: \emph{verify only}, the rate a loop sees once its sessions are up, and \emph{with session starts}, where the start of each theory's session is counted.

\paragraph{Verify only.} One worker verified 132 chunks per minute, which is 1\,607 commands per minute, and 32 workers verified 3\,181 chunks per minute, 24 times as many. The median latency of a call rose only from 0.41 to 0.52\,s, and no call timed out.

\paragraph{With session starts.} With the per-theory session starts counted, the rate at 32 workers is 899 chunks per minute, 8.0 times the single-worker 113 (8.6 times on the census-weighted measure of Table~\ref{tab:t1-series}, an efficiency of 0.27). The gap to the verify-only rate is the starts. A cold HOL-Analysis start takes 6\,s on an idle host; it took 8\,s at 8 workers, 35\,s at 16 and 156\,s at 32 (medians), with no request refused. A start loads the HOL-Analysis heap while the other
sessions check proofs. We did not investigate the cause of the slowdown
further; memory is ruled out, at 28\,GB at the peak. Acquisition is
10--14\,\% of a worker's time up to 8 workers, 30\,\% at 16 and 64\,\%
at 32. A loop whose problems share imports keeps its sessions and pays none of this; one whose problems differ in imports should expect the lower curve of Figure~\ref{fig:throughput}, or start its sessions before the work begins.

\paragraph{Reference points.} Two reference points place these rates. Against the batch tool: \code{isabelle build} checks the whole library from sources in 360\,s on this host (Table~\ref{tab:acquire}), about 27\,700 commands per minute; the replay at 32 workers checks 44\,900 commands per minute verify only and about 12\,700 with session starts. The two are not the same work, since the build is bound by the critical path of the theory dependency graph and the replay checks every theory against the prebuilt heap, but chunked, stepwise checking at 32 workers runs in the batch tool's range. Against Lean: 32 workers verify 15 chunks per second with session starts and 53 verify only, where Kimina's pool verifies about 20 whole proofs per second on 64 cores~\cite{santos2025kimina}. A chunk and a Lean proof are different units, so this is an order of magnitude, not a ranking: IPS is in that class.

\paragraph{Rollouts per hour.} Table~\ref{tab:rollouts} gives the same measurement for two scripted rollout loops. The \emph{whole-proof} loop submits eight candidates per theorem, and each candidate is one rollout. Of the eight, the corpus proof always proves, four prove when they happen to and three cannot; a candidate that stays in the document is discarded by a restore. The \emph{step-level} loop tries two alternatives at every proof command before the corpus step, each under a checkpoint (Appendix~\ref{app:bench-training} lists the candidates). About 15\,\% of the whole-proof rollouts end in a proof; that share is a property of the scripted candidates, not a result about the server. At one worker the whole-proof loop ran 6\,436 rollouts per hour and the step-level loop 2\,343 verified steps per hour; 32 workers ran 5.4 and 8.6 times as many. The time a worker spends on one rollout does not change with the number of workers: the server does not slow a worker's loop at 32. Two things bound the end-to-end rate. One is setup, nearly all of it session starts: 62--68\,\% of a worker's time at 32. The other is workers finishing early. With three theories each, one worker's loop ran 26\,s and another's 255\,s in one cell, so a fifth of the workers' time at 32 workers, and up to two fifths at 8, was spent outside any call, mostly by workers already done. Checkpoint and restore each took 3 to 4\,ms in every cell, and all 769 integrity checks of the document's source passed.

\begin{table}[!htb]
\centering\small
\setlength{\tabcolsep}{3.5pt}
\begin{tabular}{@{}r rrrrr rrrrr@{}}
\toprule
 & \multicolumn{5}{c}{Whole-proof loop} & \multicolumn{5}{c}{Step-level loop} \\
\cmidrule(lr){2-6} \cmidrule(lr){7-11}
W & rollouts/h & proved/h & s per rollout & setup & idle & paths/h & verified steps/h & s per path & setup & idle \\
\midrule
1  & 6\,436  & 954    & 0.60 & 0.22 & 0.00 & 596    & 2\,343  & 3.67 & 0.36 & 0.00 \\
8  & 20\,778 & 2\,839 & 0.60 & 0.30 & 0.39 & 1\,613 & 15\,243 & 3.65 & 0.36 & 0.19 \\
32 & 34\,518 & 4\,798 & 0.60 & 0.68 & 0.21 & 2\,071 & 20\,042 & 3.67 & 0.62 & 0.23 \\
\bottomrule
\end{tabular}
\caption{Scripted rollout loops at $W$ workers, two repeats each. A path is one step-level rollout: a theorem's proof walked command by command. Rates
  are end to end: session acquisition, entering the theory and replaying
  up to each problem included. ``s per rollout'' is a worker's loop time
  per rollout on the theories every cell shares, the like-for-like
  comparison across $W$. Setup is the share of a worker's time in
  acquisition, entering and replaying to the problems; idle the share of
  the workers' time outside any call, mostly workers already finished. No
  rollout ended in a memory,
  prover-death or session error; the candidates' timeout share was
  0.5--2\,\%.}
\label{tab:rollouts}
\end{table}

\begin{figure}[!htb]
\centering
\input{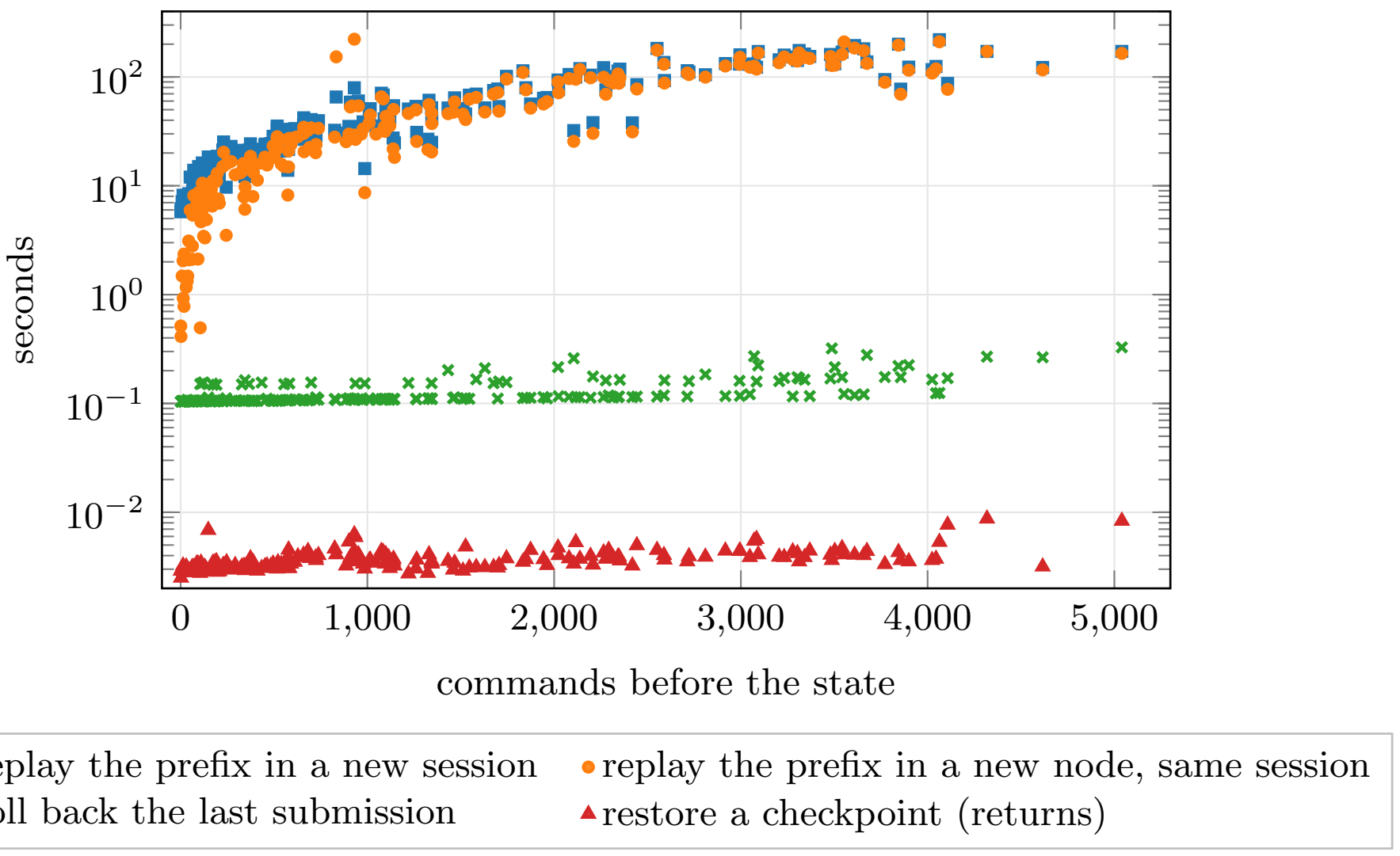}
\caption{Time to go back to a proof state, against the number of commands
  before it, for 200 states: by a restore call, by a rollback, and by
  replaying the prefix in a new theory node of the same session or in a
  new session. At this scale the branch cost, the restore call plus its
  settling delay, cannot be told from the restore call alone;
  Table~\ref{tab:branch} gives it.}
\label{fig:branch}
\end{figure}

\paragraph{Branch cost.} We took 200 proof states across 68 theories,
a third each early, in the middle and late in their files. At each we
tried the next block and then went back to the state in three ways: by
restoring a checkpoint, by rolling back the last submission, and by
replaying the theory up to the state in a new theory node or a new
session. The replay is the cost of a branch when no state was saved
(Figure~\ref{fig:branch}, Table~\ref{tab:branch}). A restore call returns in 4\,ms, before the document has settled, so the
verification call that follows may have to wait. We call that extra time
the \emph{settling delay}: the time by which a verification call is
slower after a restore than on a settled document. The branch cost, the
restore call plus its settling delay, is 4\,ms at the median and 20\,ms
on average (200\,ms at the 95th percentile); for the late states its
mean is 25\,ms, with a 95\,\% confidence bound of 63\,ms.
Replaying to the same state takes 8\,s at the median for the early
states, 43\,s in the middle and 91\,s late in a file, 13.9, 49.6 and
93.2\,s with a fresh session, so the branch cost is at least 350 times lower
than a replay's early in a file and 1\,450 times late in it.
Verified again after each restore and each rollback, the block gave the
same report as the first time in all 1\,000 repeats, and the source
matched after all 600 restores and 400 rollbacks.

\begin{table}[!htb]
\centering\small
\begin{tabular}{@{}l rr@{}}
\toprule
Operation & p50 & p95 \\
\midrule
Save a checkpoint & 3\,ms & 4\,ms \\
Restore a checkpoint (the call) & 4\,ms & 5\,ms \\
Branch cost: restore call plus settling delay & 4\,ms & 200\,ms \\
Roll back the last submission (settled) & 110\,ms & 216\,ms \\
Replay the prefix in a new node, same session & 30\,s & 165\,s \\
Replay the prefix in a new session & 35\,s & 169\,s \\
\bottomrule
\end{tabular}
\caption{The cost of going back to a proof state, over 200 states. The
  replay rows pool the three positions; Figure~\ref{fig:branch} shows
  them against the prefix length.}
\label{tab:branch}
\end{table}

\subsection{Checks}
\label{sec:checks}

Two properties of the implementation are checked here rather than
measured, since they are things a service must get right rather than
metrics on which to compare services: that each verdict flag catches the
failure it exists to catch, and that a looping step cannot take a worker
with it. The tables are in Appendix~\ref{app:bench-training}.

\paragraph{Verdict flags.} We tested each flag against the failure it exists to
catch, with a batch build of the same text as the reference
(Table~\ref{tab:agreement}). We sampled 200 theorem blocks across the
library and spliced each into its theory in four forms: unchanged, the proof
replaced by \code{sorry}, the last closing step removed, and the first
fact name made undefined. Every flag caught every case of its failure:
\code{used\_sorry} the 200 placeholders, \code{proof\_open} the 200
unfinished proofs, \code{success} the 197 failing steps, 13 of them by
the budget, since the step ran until its bound and was rolled back rather than
failing at once. The 200 unchanged blocks and the 93 theories as written
were accepted by the flags and by the build alike. No case was proved
in-session and rejected by the build, the disagreement that would corrupt
a reward, and no mutated form was accepted by either. The 890 checks,
the five rows of Table~\ref{tab:agreement}, rest on 690
batch builds, the 93 theories as written and the 597 mutated forms,
since an unchanged block shares the build of its theory as written; the
builds took 13.6\,s at the median and 24\,s at p95, four at a time. A
loop can therefore read its reward from the flags and run the batch build
where a verdict is final; Section~\ref{sec:verdict} says what no flag
sees.

\paragraph{Bounded step.} We placed 100 steps meant to loop before proved
theorems of 9 theories, 50 of each kind: a rewriting loop and a
\code{metis} call on a false goal (Appendix~\ref{app:bench-training}). Each was submitted
under a budget of 5\,s, and some of those that looped were tried again under 30
and
120\,s (Table~\ref{tab:bounded}). Every looping step returned at its
budget with the stuck command named, the server rolled it back within 0.6\,s at
p95 in every row but one (4.2\,s), the document's source was unchanged
after every attempt, and the next corpus block proved at the census's
speed (0.41\,s median). Twelve of the rewriting steps failed at
once and are counted apart. At 8 workers, with 15 looping steps of 120\,s
spread over the workers so that up to seven loops ran at once, the other
workers' blocks ran at the speed they had in the 8-worker throughput cell,
where nothing loops (median ratio 1.00 over 1\,130 matched blocks), and
waited 4--5\,ms for the server. A long rewriting loop does leave memory in the prover:
1.2\,GB at the median after a 30\,s loop and 7\,GB after 120\,s, which the
rollback does not reclaim; the \code{metis} loops left 130\,MB or less at the median (858\,MB at most). This
is Poly/ML's own memory, which grows under a long rewrite and is not returned
to the operating system afterwards. A cap on each prover's memory
(Appendix~\ref{app:bench}) bounds it; a prover that reaches the
cap fails with an out-of-memory error that the response names as such, and the
memory gate sees the growth on the next call.

\section{Multi-agent serving}
\label{sec:serving}

This section covers the \textbf{multi-agent serving scenario}: how the pool behaves on a
shared host, then the measurements against the requirements of
Section~\ref{sec:serving-req}.

\subsection{Design}

Section~\ref{sec:acquire} describes the pool as a client meets it; three
things concern the host. Admission is gated on the container's memory:
above the threshold the pool evicts the least-recently-used idle
session, then refuses a new session with an explanation, so the host
reports ``busy'' instead of suffering an out-of-memory kill. Concurrent
Sledgehammer runs are capped by a server-wide limit, since each fans out
to several external provers. All sessions share one gateway JVM; its
crash is detected and the pool rebuilt. Probes, metrics and the lack of
authentication are in Appendix~\ref{app:deploy}.

\subsection{Results}
\label{sec:serving-measured}

All numbers in this section come from one run on a 32-core, 128\,GB
Ubuntu host with Isabelle2026, both tools native and nothing else
running; Appendix~\ref{app:bench} gives the method and the workloads,
the results repository the per-cell results. We write $k$ for the
number of concurrent sessions. We took the capacity and isolation measurements for IPS and,
under identical conditions, for PIDE MCP at commit \code{dd46315}
(19 October 2026). The figure and the table of this section show
PIDE MCP in two arrangements: \emph{PIDE MCP A}, one PIDE MCP process
holding $k$ sessions, and \emph{PIDE MCP B}, $k$ processes with one
session each (Section~\ref{sec:comparison-measured} says why both).
This section reports the IPS numbers; Section~\ref{sec:comparison}
reads the two side by side.

\paragraph{Memory measures.} We report two measures. The
\emph{proportional set size} (PSS) of the tool's processes, summed,
counts a shared page once. \emph{Host used} is the change in
\code{MemTotal} minus \code{MemAvailable} over the cell's baseline; it
checks that no memory went uncounted. A prover shares only about
13\,MB, so its PSS, resident and unique sizes agree within 2\,\%. Under a memory limit we also read the
cgroup's counters, which the limit is enforced on
(Appendix~\ref{app:bench}).

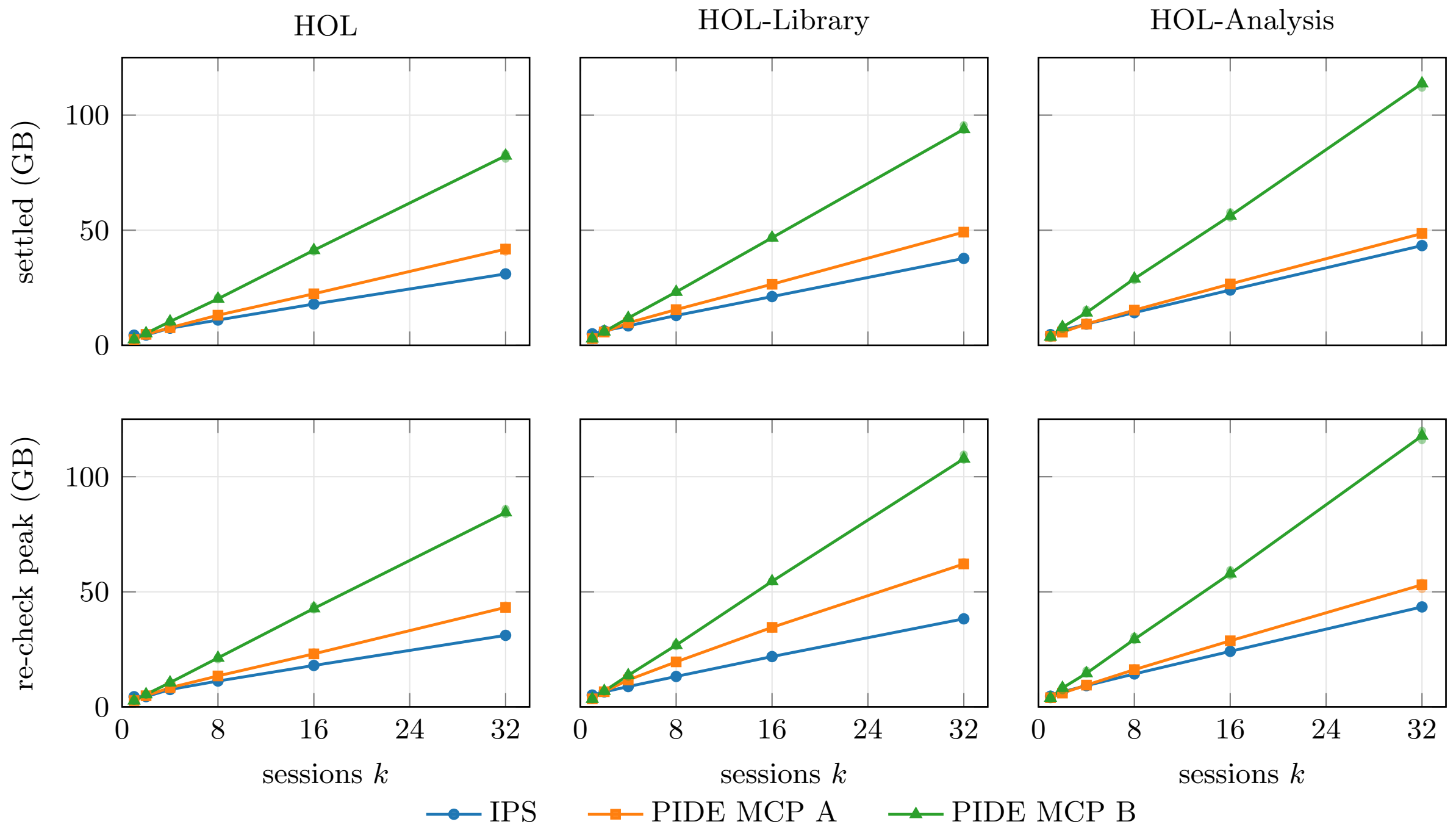
\begin{figure}[!htb]
\centering
\begin{tikzpicture}
\begin{groupplot}[report,
  group style={group size=3 by 2, horizontal sep=0.55cm, vertical sep=0.8cm,
    xlabels at=edge bottom, xticklabels at=edge bottom,
    ylabels at=edge left, yticklabels at=edge left},
  width=6.0cm, height=4.7cm, xmin=0, xmax=34, ymin=0, ymax=125,
  xtick={0,8,16,24,32}, xlabel={sessions $k$},
  title style={font=\footnotesize, yshift=-3pt}]
\nextgroupplot[title={HOL}, ylabel={settled (GB)}, legend to name=memlegend, legend columns=3, legend style={draw=none, fill=none, /tikz/every even column/.append style={column sep=1.2em}}]
\addplot[plotblue, mark=*, mark size=1.3pt] coordinates {(1,4.42) (2,4.47) (4,7.45) (8,11.00) (16,17.93) (32,31.03)};
\addlegendentry{IPS}
\addplot[repeats, plotblue] coordinates {(1,4.39) (1,4.44) (1,4.42) (2,5.12) (2,4.47) (2,4.40) (4,7.44) (4,7.45) (4,7.45) (8,11.08) (8,10.89) (8,11.00) (16,17.83) (16,17.93) (16,17.97) (32,31.17) (32,31.03) (32,30.97)};
\addplot[plotorange, mark=square*, mark size=1.2pt] coordinates {(1,2.57) (2,4.76) (4,7.66) (8,13.10) (16,22.39) (32,41.80)};
\addlegendentry{PIDE MCP A}
\addplot[repeats, plotorange] coordinates {(1,2.48) (1,2.77) (1,2.57) (2,4.76) (2,4.68) (2,4.94) (4,7.66) (4,7.45) (4,8.32) (8,12.88) (8,13.15) (8,13.10) (16,22.44) (16,22.39) (16,22.12) (32,41.80) (32,40.71) (32,42.47)};
\addplot[plotgreen, mark=triangle*, mark size=1.5pt] coordinates {(1,2.56) (2,5.19) (4,10.33) (8,20.20) (16,41.30) (32,82.34)};
\addlegendentry{PIDE MCP B}
\addplot[repeats, plotgreen] coordinates {(1,2.47) (1,2.56) (1,2.91) (2,5.08) (2,5.37) (2,5.19) (4,10.33) (4,10.52) (4,9.87) (8,20.16) (8,20.20) (8,20.83) (16,40.76) (16,41.60) (16,41.30) (32,83.17) (32,82.34) (32,81.07)};
\nextgroupplot[title={HOL-Library}]
\addplot[plotblue, mark=*, mark size=1.3pt] coordinates {(1,5.01) (2,6.32) (4,8.45) (8,12.96) (16,21.18) (32,37.75)};
\addplot[repeats, plotblue] coordinates {(1,5.01) (1,5.05) (1,4.91) (2,6.18) (2,6.46) (2,6.32) (4,8.45) (4,8.54) (4,8.39) (8,12.96) (8,12.86) (8,12.97) (16,21.12) (16,21.18) (16,21.29) (32,37.78) (32,37.75) (32,37.70)};
\addplot[plotorange, mark=square*, mark size=1.2pt] coordinates {(1,2.91) (2,5.79) (4,9.76) (8,15.52) (16,26.55) (32,49.17)};
\addplot[repeats, plotorange] coordinates {(1,3.04) (1,2.91) (1,2.89) (2,6.03) (2,5.79) (2,5.47) (4,9.90) (4,9.76) (4,9.72) (8,15.90) (8,15.52) (8,15.43) (16,27.12) (16,26.31) (16,26.55) (32,49.07) (32,49.68) (32,49.17)};
\addplot[plotgreen, mark=triangle*, mark size=1.5pt] coordinates {(1,2.92) (2,5.94) (4,11.86) (8,23.19) (16,46.73) (32,93.87)};
\addplot[repeats, plotgreen] coordinates {(1,2.92) (1,2.90) (1,3.10) (2,5.94) (2,5.89) (2,5.94) (4,11.76) (4,11.97) (4,11.86) (8,23.19) (8,23.17) (8,23.34) (16,46.73) (16,46.66) (16,47.21) (32,93.87) (32,93.58) (32,95.63)};
\nextgroupplot[title={HOL-Analysis}]
\addplot[plotblue, mark=*, mark size=1.3pt] coordinates {(1,4.67) (2,6.54) (4,9.24) (8,14.22) (16,24.01) (32,43.32)};
\addplot[repeats, plotblue] coordinates {(1,4.40) (1,4.67) (1,5.05) (2,6.54) (2,6.50) (2,6.64) (4,9.08) (4,9.28) (4,9.24) (8,14.22) (8,14.18) (8,14.27) (16,23.95) (16,24.01) (16,24.01) (32,43.32) (32,43.24) (32,43.38)};
\addplot[plotorange, mark=square*, mark size=1.2pt] coordinates {(1,3.98) (2,5.75) (4,9.25) (8,15.29) (16,26.65) (32,48.58)};
\addplot[repeats, plotorange] coordinates {(1,3.48) (1,3.98) (1,3.98) (2,5.89) (2,5.46) (2,5.75) (4,9.26) (4,9.25) (4,9.19) (8,15.38) (8,15.29) (8,15.22) (16,26.69) (16,26.65) (16,26.40) (32,48.80) (32,47.82) (32,48.58)};
\addplot[plotgreen, mark=triangle*, mark size=1.5pt] coordinates {(1,3.71) (2,7.92) (4,14.19) (8,28.95) (16,56.28) (32,113.72)};
\addplot[repeats, plotgreen] coordinates {(1,3.19) (1,3.71) (1,4.01) (2,7.92) (2,8.21) (2,7.04) (4,14.19) (4,14.09) (4,15.49) (8,29.08) (8,28.95) (8,28.34) (16,56.28) (16,57.79) (16,55.49) (32,113.72) (32,113.75) (32,111.91)};
\nextgroupplot[ylabel={re-check peak (GB)}]
\addplot[plotblue, mark=*, mark size=1.3pt] coordinates {(1,4.45) (2,4.48) (4,7.57) (8,11.25) (16,18.02) (32,31.10)};
\addplot[repeats, plotblue] coordinates {(1,4.41) (1,4.46) (1,4.45) (2,5.13) (2,4.48) (2,4.41) (4,7.57) (4,7.82) (4,7.48) (8,11.26) (8,11.25) (8,11.21) (16,17.93) (16,18.02) (16,18.04) (32,31.26) (32,31.10) (32,31.10)};
\addplot[plotorange, mark=square*, mark size=1.2pt] coordinates {(1,2.77) (2,4.86) (4,8.37) (8,13.47) (16,23.06) (32,43.25)};
\addplot[repeats, plotorange] coordinates {(1,2.77) (1,2.82) (1,2.67) (2,4.78) (2,4.86) (2,4.96) (4,8.37) (4,8.08) (4,8.63) (8,13.91) (8,13.41) (8,13.47) (16,23.06) (16,23.74) (16,22.84) (32,43.25) (32,43.15) (32,44.05)};
\addplot[plotgreen, mark=triangle*, mark size=1.5pt] coordinates {(1,2.71) (2,5.45) (4,10.57) (8,21.30) (16,42.78) (32,84.47)};
\addplot[repeats, plotgreen] coordinates {(1,2.66) (1,2.71) (1,3.17) (2,5.45) (2,5.54) (2,5.31) (4,10.57) (4,10.80) (4,10.40) (8,20.69) (8,21.30) (8,21.37) (16,42.34) (16,42.78) (16,43.78) (32,85.99) (32,84.47) (32,83.77)};
\nextgroupplot[]
\addplot[plotblue, mark=*, mark size=1.3pt] coordinates {(1,5.16) (2,6.55) (4,8.85) (8,13.24) (16,21.88) (32,38.28)};
\addplot[repeats, plotblue] coordinates {(1,5.16) (1,5.17) (1,5.04) (2,6.40) (2,6.58) (2,6.55) (4,8.85) (4,9.03) (4,8.65) (8,13.28) (8,13.14) (8,13.24) (16,21.88) (16,21.77) (16,21.93) (32,38.61) (32,38.28) (32,38.13)};
\addplot[plotorange, mark=square*, mark size=1.2pt] coordinates {(1,3.53) (2,6.61) (4,11.73) (8,19.58) (16,34.56) (32,62.10)};
\addplot[repeats, plotorange] coordinates {(1,3.53) (1,3.56) (1,3.52) (2,6.85) (2,6.61) (2,6.55) (4,11.45) (4,11.76) (4,11.73) (8,20.07) (8,19.48) (8,19.58) (16,34.92) (16,33.90) (16,34.56) (32,62.08) (32,62.98) (32,62.10)};
\addplot[plotgreen, mark=triangle*, mark size=1.5pt] coordinates {(1,3.44) (2,6.89) (4,13.74) (8,26.84) (16,54.54) (32,107.80)};
\addplot[repeats, plotgreen] coordinates {(1,3.44) (1,3.30) (1,3.61) (2,6.72) (2,6.98) (2,6.89) (4,13.74) (4,13.92) (4,13.14) (8,27.49) (8,26.84) (8,26.42) (16,54.52) (16,54.68) (16,54.54) (32,107.80) (32,107.40) (32,109.62)};
\nextgroupplot[]
\addplot[plotblue, mark=*, mark size=1.3pt] coordinates {(1,4.67) (2,6.54) (4,9.26) (8,14.34) (16,24.14) (32,43.45)};
\addplot[repeats, plotblue] coordinates {(1,4.41) (1,4.67) (1,5.05) (2,6.54) (2,6.53) (2,6.68) (4,9.08) (4,9.28) (4,9.26) (8,14.34) (8,14.26) (8,14.45) (16,24.14) (16,24.16) (16,24.14) (32,43.45) (32,43.38) (32,43.49)};
\addplot[plotorange, mark=square*, mark size=1.2pt] coordinates {(1,4.02) (2,6.02) (4,9.49) (8,16.21) (16,28.73) (32,53.06)};
\addplot[repeats, plotorange] coordinates {(1,3.65) (1,4.15) (1,4.02) (2,6.29) (2,5.67) (2,6.02) (4,9.49) (4,9.62) (4,9.28) (8,16.86) (8,16.21) (8,15.94) (16,28.74) (16,28.61) (16,28.73) (32,53.91) (32,51.19) (32,53.06)};
\addplot[plotgreen, mark=triangle*, mark size=1.5pt] coordinates {(1,3.88) (2,8.23) (4,14.59) (8,29.37) (16,57.89) (32,117.76)};
\addplot[repeats, plotgreen] coordinates {(1,3.33) (1,3.88) (1,4.12) (2,8.23) (2,8.54) (2,7.25) (4,14.59) (4,14.52) (4,15.77) (8,30.59) (8,29.37) (8,29.06) (16,57.89) (16,59.41) (16,57.10) (32,119.93) (32,117.76) (32,115.90)};
\end{groupplot}
\path (group c1r2.south west) -- (group c3r2.south east)
  node[midway, below=0.75cm] {\pgfplotslegendfromname{memlegend}};
\end{tikzpicture}
\caption{Memory of the tool's processes (summed PSS) against the number
  of sessions $k$, one column per heap, three repeats per point as small
  marks with the line through their median. Top: after every session has
  loaded and checked its theory. Bottom: at the concurrent re-check of
  all $k$ theories (the re-check peak, Appendix~\ref{app:bench}). IPS: one pool.
  PIDE MCP A: one process holding $k$ sessions. PIDE MCP B: $k$ processes,
  one session each.}
\label{fig:capacity}
\end{figure}

\paragraph{Capacity.} Figure~\ref{fig:capacity} plots the summed PSS of
the tool's processes against the number of concurrent sessions for three
heaps, each session holding a checked theory, three repeats per point. The curves are straight to
$k$\,=\,32. A least-squares fit on the processes' summed PSS
(Table~\ref{tab:capacity}) separates the pool's fixed cost, one server
process and one gateway JVM, 3.7--4.2\,GB, from the cost of one more
session: 886\,MB on HOL, 1\,077\,MB on HOL-Library and 1\,263\,MB on
HOL-Analysis, which is the prover alone. The gateway JVM grows from
3.6\,GB at one session to 5.1\,GB at 32, so the fixed cost is not
constant, but its growth is small against the sessions. After the fixed
cost, a 100\,GB budget (102\,400\,MB) holds 111 HOL, 91 HOL-Library or
77 HOL-Analysis sessions. The same fit on host used gives per-session
costs within 1\,\% of these, so the process view accounts for everything
the host loses. These are the \emph{settled} costs of a session holding one
checked theory, a planning estimate; a long proof search grows its
prover beyond them (Table~\ref{tab:bounded}). When all $k$ sessions
re-check their theories at once, after a line is inserted near the top
(the lower row of Figure~\ref{fig:capacity}, Appendix~\ref{app:bench}),
the per-session cost rises by at most 1.4\,\% under IPS, because the re-check
runs inside the memory each prover already holds; under PIDE MCP~A it rises by
4--28\,\%, most on HOL-Library.

\begin{table}[!htb]
\centering\small
\setlength{\tabcolsep}{4pt}
\begin{tabular}{@{}llrrrr@{}}
\toprule
Tool & Heap & \shortstack{Settled MB\\per session} & \shortstack{Re-check peak\\MB per session} & \shortstack{Sessions\\per\\100\,GB} & \shortstack{Cold start\\p50/p95 s} \\
\midrule
IPS & HOL & 886 & 887 & 111 & 2.3 / 3.9 \\
IPS & HOL-Library & 1077 & 1092 & 91 & 3.4 / 5.1 \\
IPS & HOL-Analysis & 1263 & 1267 & 77 & 6.0 / 8.2 \\
\midrule
PIDE MCP A & HOL & 1266 & 1320 & 78 & 2.0 / 3.2 \\
PIDE MCP A & HOL-Library & 1495 & 1915 & 66 & 3.0 / 4.3 \\
PIDE MCP A & HOL-Analysis & 1461 & 1602 & 67 & 5.8 / 7.1 \\
\midrule
PIDE MCP B & HOL & 2630 & 2712 & 38 & 3.1 / 3.2 \\
PIDE MCP B & HOL-Library & 3017 & 3466 & 33 & 4.1 / 4.2 \\
PIDE MCP B & HOL-Analysis & 3604 & 3753 & 28 & 6.7 / 7.3 \\
\bottomrule
\end{tabular}
\caption{Capacity and latency from the capacity runs (E1 in
  Appendix~\ref{app:bench}), on the processes'
  summed PSS. Settled MB per session: the least-squares marginal cost over
  $k$\,=\,1\ldots32 (18 cells per row) after the theories are loaded; peak: the same at
  the concurrent re-check; sessions in a 100\,GB budget (102\,400\,MB)
  after the fit's fixed cost, rounded down; the cold-start latency.
  The fixed cost is the fit's intercept: 3.7--4.2\,GB for IPS (the server
  and the gateway JVM) and 2.4--3.2\,GB for PIDE MCP A (its JVM); PIDE
  MCP B has no fixed process and its intercepts lie within 0.5\,GB of
  zero.}
\label{tab:capacity}
\end{table}

\paragraph{Acquisition latency.} Table~\ref{tab:acquire} separates the three
cases. A warm hit, a released session handed to the next client with the
same imports, takes 2\,ms and returned the same session in all 36 trials.
A cold start on a built heap takes 2 to 6 seconds depending on the heap.
A heap build takes minutes: 114\,s for HOL-Library and 360\,s for
HOL-Analysis from the distribution sources on this host. A cold start
costs over a thousand times a warm hit, and a heap build 30 to 60 times
a cold start, which is why IPS keeps sessions warm and keeps a built
heap until its sources change.

\begin{table}[!htb]
\centering\small
\begin{tabular}{@{}lrrr@{}}
\toprule
 & HOL & HOL-Library & HOL-Analysis \\
\midrule
Warm hit (a released session acquired again), p50 & 0.002\,s & 0.002\,s & 0.002\,s \\
Cold start on a built heap, p50 / p95 & 2.3 / 3.9\,s & 3.4 / 5.1\,s & 6.0 / 8.2\,s \\
Heap build from sources (once) & 4\,s$^{\dagger}$ & 114\,s & 360\,s \\
\bottomrule
\end{tabular}
\caption{Acquisition latency under IPS, by case. Warm hits: 4 sessions, 3
  rounds, every acquisition reused the session its client had released. Cold starts:
  the session starts of the capacity runs. Heap builds: the base heaps on this host.
  $^{\dagger}$HOL was already built by the Isabelle installation; the
  figure is the check, not a build.}
\label{tab:acquire}
\end{table}

\paragraph{Pressure.} We started the server in a systemd scope with
\code{MemoryMax=48G} and added HOL-Analysis sessions one at a time, each
loading and checking the theory, until the pool refused. IPS admitted 27
sessions and refused the 28th with HTTP 503 and a body that states the
facts: 37.5\,GB in use (78\,\% of the limit), 10.5\,GB available, every
running session leased and so none evictable, retry in 5\,s. The gate
refused because the next session's 4\,GB estimate would cross the
85\,\% threshold. No
process died (\code{oom\_kill} stayed 0; the scope's peak was 37.5\,GB).
After the client released half of its sessions, the next session was
admitted and loaded 8.4\,s later. The stop is the gate's margin, not the
limit: these runs used the default 4\,GB estimate per HOL-Analysis
session, and the limit itself would hold about 35 sessions of the
measured 1.2\,GB. So the 27 is where the gate stops a client, not how
many sessions the memory holds.

\paragraph{Isolation.} With 8 HOL sessions leased, we killed the newest
session's prover process with SIGKILL and asked every session to re-check
its theory within 30\,s. The other 7 answered. The victim's session
returned ``prover process died, the session has been discarded'', and
its slot went back to the pool. The gateway and the server process were
unaffected. A prover death is thus one client's failure, reported as
such. Cross-lease requests, the other half of isolation, are refused by
the lease check; we did not measure them here.

\section{Comparison with PIDE MCP}
\label{sec:comparison}

IPS and PIDE MCP~\cite{kappelmann2026pidemcp} are both direct PIDE
frontends (Section~\ref{sec:position}), and both expose their sessions
to a language model over MCP. This section says which version of PIDE
MCP we measured, where the two differed at that version and which of the
differences are tooling, what we measured for both under the same
conditions, and what each offered that the other did not. The full
inventory, 24 concerns side by side, is
\path{docs/FEATURE_INVENTORY.md} in the repository.

\subsection{The version measured}
\label{sec:pide-version}

Every statement about PIDE MCP in this report is about commit
\code{dd46315} of 19 October 2026 (server 0.1.0), five weeks after the
preprint, unless it says otherwise. At that commit PIDE MCP was a
headless, multi-session PIDE
frontend with a replaceable tool set, serving one client over MCP on
standard input and output. Three things have changed since, and each
bears on a result below.
\begin{itemize}
  \item \emph{Name and transport.} The project was renamed PIDE AI on
    4 November 2026. Since its Isabelle2026 release a server also
    listens on a TCP port, so several agents can share one server and
    its sessions. Topology B of Section~\ref{sec:comparison-measured},
    one process per agent, was how independent agents had to use the
    commit we measured; it is no longer needed.
  \item \emph{Waiting.} Since 31 October 2026 \code{get\_state} takes a
    timeout and waits until the commands have finished or failed, and
    \code{edit\_get\_state} joins the edit to that wait. The polling
    loop of our harness (Section~\ref{sec:comparison-measured}) is not
    needed against these versions.
  \item \emph{Visibility.} Ranges that an agent reads or edits are no
    longer made visible by default (commit \code{8fdfc12}, 14 November
    2026), a change made after we reported to its author the memory cost of
    Section~\ref{sec:comparison-measured}.
\end{itemize}
We have repeated no measurement on a later version. The preprint says
that the core tools can be removed like any others and that ``the
concrete set of core tools is likely to change over
time''~\cite{kappelmann2026pidemcp}. What we measured is the tool set of
one commit, and we say so wherever a difference could be read as one of
architecture.

\subsection{Differences at the measured commit}

\paragraph{Where a node's text comes from.} Both tools edit the nodes of
a PIDE document. In IPS a session has one theory node whose text arrives
in requests: the server appends to it or replaces it, no file stands
behind it, and the session's imports are fixed when it starts. In PIDE
MCP the nodes are those of a project, theories and auxiliary files such
as ML files, each theory with its own header. PIDE loads them from disk
and then manages them in memory, and PIDE MCP's \code{read} and
\code{edit} tools synchronise a node's content with its file at every
call~\cite{kappelmann2026pidemcp}. That gives PIDE MCP project scope,
dependency resolution, positioned messages, and co-presence with a
human's editor through the file system. IPS rolls back a chunk as a unit when it fails,
records checkpoints and restores them; PIDE MCP had no tool that
undoes an edit at that commit. Such a tool could be written: a PIDE
session keeps a history of its document's
versions, and an edit tool could keep its modifications on a stack.

\paragraph{Waiting.} The preprint says that PIDE MCP ``decidedly does
not block on checking and does not use timeouts for its core tools, but
users can implement such means for their own tools if
desired''~\cite{kappelmann2026pidemcp}. At the commit we measured an
edit returns when PIDE has accepted it, the agent reads a snapshot with
\code{get\_state} and, as the preprint notes, typically polls until the
result is there, and a long-running tool call can be cancelled through
the protocol. An IPS call waits for its submission to settle under a
budget, and a submission that exceeds the budget is rolled back and reported
as a timeout. The first suits an agent that can look again later; the
second suits a loop with a fixed worker pool that cannot afford to lose
a worker. A tool that waits
under a bound can be written for PIDE MCP, and its later versions ship
one (Section~\ref{sec:pide-version}).

\paragraph{Decisions across sessions.} IPS has a pool with
leases, idle and lease timeouts, LRU eviction, a memory admission
gate and a Sledgehammer cap. PIDE MCP, at the commit we measured,
reports session statistics and leaves admission and reclamation to the
deployment, a reasonable division of labour for one agent on one
machine. For many agents on one host the deployment has to supply
them.

\paragraph{Architecture or tooling.} In the preprint's sense the two
tools have one architecture: each is a PIDE frontend that manages
several headless sessions through the Isabelle/Scala
API~\cite{kappelmann2026pidemcp}. None of the differences above is
architectural in that sense. One of them is a decision about what a
session holds, and IPS could not change it without changing its
sessions. A session's context is fixed and its one theory has
no file behind it, so sessions with the same imports can stand in for one
another, which is what lets the pool reuse them. In exchange a session
cannot hold a project of several theories
(Section~\ref{sec:limitations}).
Everything else in this section is tooling at one commit, and it
transfers in both directions:
budgets, rollback on failure, checkpoints and the verdict flags could be
written as PIDE MCP tools, and positioned messages, cancellation and
progress could be added to IPS. The lease belongs with the tooling too.
No verification feature needs it; it is what tells the layer above the
sessions which of them are in use, so that the layer evicts only the others.

\subsection{Measurements}
\label{sec:comparison-measured}

We ran the capacity and isolation experiments of
Section~\ref{sec:serving-measured} for PIDE
MCP at commit \code{dd46315} on the same host, Isabelle build,
prover memory cap and workloads, in two arrangements. \emph{Topology A} is
one PIDE MCP process holding $k$ sessions for one client, the
multi-session use the tool was built for. \emph{Topology B} gives every
agent its own PIDE MCP process, which independent agents needed at that
commit (Section~\ref{sec:pide-version}). It is a cost study of that
arrangement and not a defect of the tool.

\paragraph{Memory per session.} Under topology A a session costs 0.2 to
0.4\,GB more than under IPS on every heap, and 100\,GB holds 66--78
sessions rather than 77--111 (Table~\ref{tab:capacity}). The gap is inside the prover
processes, not the JVMs: over the cells of the series a PIDE MCP~A
prover holds 0.94--1.25\,GB on HOL and 1.29--1.39\,GB on HOL-Analysis
(medians per cell), an IPS prover 0.79 and 1.17\,GB, while the two
tools' JVMs end within 0.4\,GB of each other at 32 sessions. The likely
cause is the PIDE perspective. At this commit PIDE MCP's \code{edit}
marks the edited range and 50 lines around it as visible (option
\path{pide_mcp_range_context}). Our workload writes the whole theory in
one edit, so the whole theory is visible and PIDE runs its print
functions on every command, among them Isabelle's automatic tools, to
which PIDE AI's change log attributes the cost. IPS gives its nodes an
empty visible range (Appendix~\ref{sec:pide-terms}). We reproduced the
gap on IPS by making its nodes fully visible: the same HOL theory then
cost 0.98 to 1.22\,GB per prover in seven of eight provers, against
0.80\,GB with the empty range (Appendix~\ref{app:bench} gives the check,
Section~\ref{sec:limitations} what it does not cover). If the perspective is the cause, the extra memory is the cost of those print functions, and it
depends on the workload. At this commit only the range last read or
edited stays visible, with its 50 lines of context. We therefore expect an
agent that edits a few lines at a time to pay little of it, though we
did not measure that case. The session options do not explain the gap:
IPS ran \code{parallel\_proofs=2} and PIDE MCP Isabelle's default of 1,
which give the same checking in a headless session
(Appendix~\ref{sec:pide-terms}).

PIDE MCP's fixed cost is lower (2.4--3.2\,GB against
3.7--4.2\,GB): its JVM starts at 1.5\,GB on HOL and grows to 4.9\,GB at
32 sessions, where the IPS gateway starts at 3.6\,GB and ends at
5.1\,GB. So PIDE MCP A is the smaller deployment for a few sessions and
IPS for many: the totals cross between three and five sessions,
depending on the heap. Under topology B a
session costs 2.6--3.5\,GB, three times IPS, because every agent pays a
1.5--2.2\,GB JVM; 32 HOL-Analysis agents used 114\,GB of the 128\,GB
host, 118\,GB at the re-check peak.

\paragraph{Throughput.} We ran the training workload of
Section~\ref{sec:training-measured} under both tools, with the same
theories, census and heap, and each theory on a fresh session started
inside the timed window. For PIDE MCP our harness used an
edit-and-poll loop over its tools: one \code{edit} per block, then
\code{get\_state} every 0.1\,s until the block has settled. At one
worker the two were level: IPS verified 132 chunks per minute and the
PIDE MCP loop 135, or 113 and 116 with the session starts counted. At
eight workers IPS verified 927 chunks per minute against 813, and 497
against 381 with the starts. The loop and its interval of 0.1\,s
are ours and not PIDE MCP's: we chose the interval because IPS re-reads
the node at it. The two calls do not do the same work. \code{verify\_chunk} does more
than check the chunk: when the chunk fails it rolls the chunk back, and
when the chunk succeeds it asks the prover whether a proof is still
open, which costs about one 0.1\,s poll. In both cases it returns a
verdict with the status of each command. The PIDE MCP loop returns
counts of commands, from which the client reads off a verdict itself. Both rates count verified chunks, so a block that
fails or times out lowers them rather than inflating them. Some blocks fail under the PIDE MCP loop through choices of our driver
and not of the tool: one block per cell that did not settle and ended
its theory, and under 1\,\% of the blocks through the driver's file
name and its closing \code{end} (Appendix~\ref{app:bench-training}).

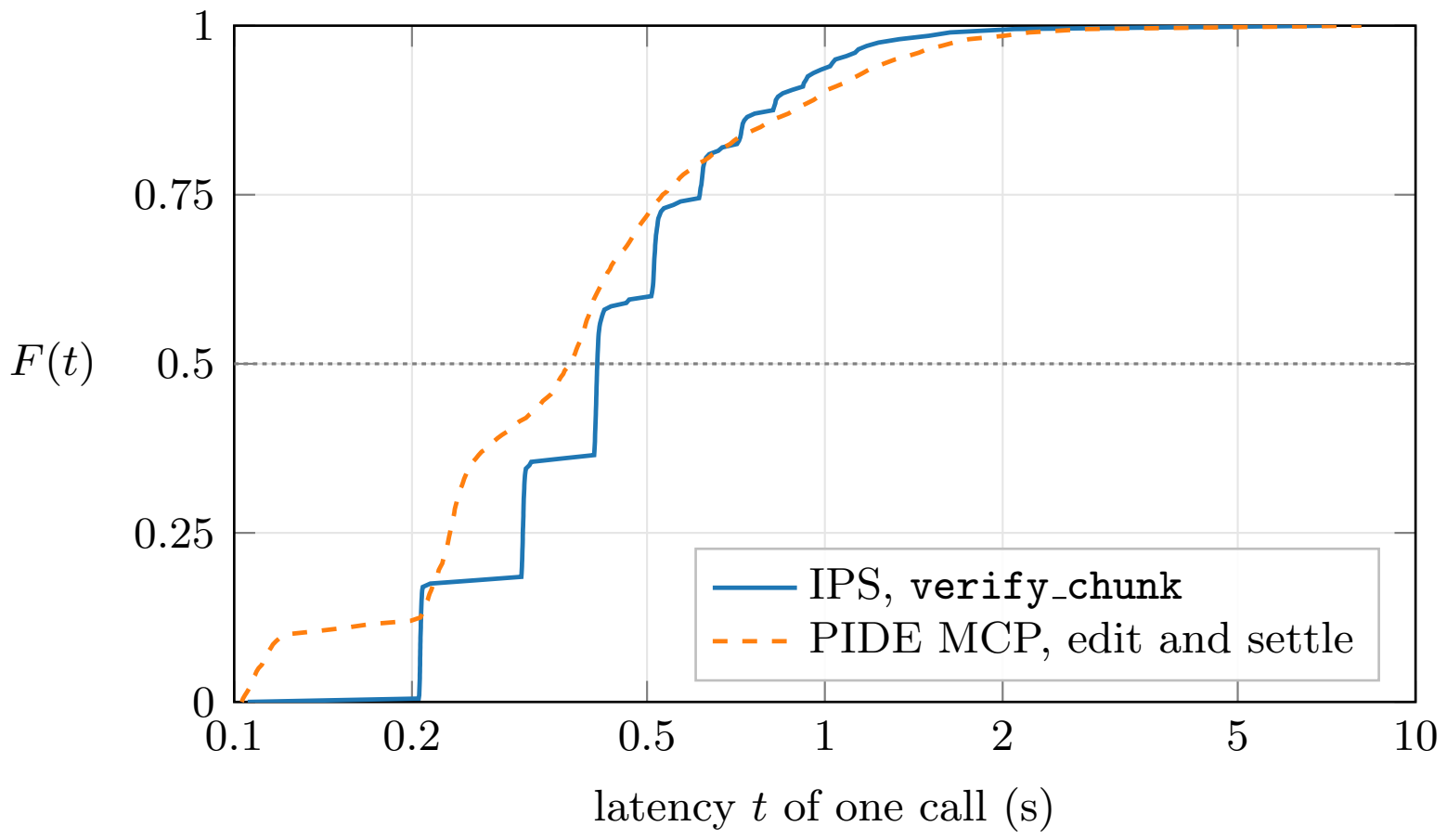
\begin{figure}[!htb]
\centering
\begin{tikzpicture}
\begin{axis}[report, width=10cm, height=6.4cm,
  xmode=log, log ticks with fixed point,
  xmin=0.1, xmax=10, ymin=0, ymax=1,
  xtick={0.1,0.2,0.5,1,2,5,10},
  ytick={0,0.25,0.5,0.75,1},
  xlabel={latency $t$ of one call (s)}, ylabel={$F(t)$}, ylabel style={rotate=-90},
  legend pos=south east]
\addplot[gray, densely dotted, line width=0.6pt, forget plot] coordinates {(0.1,0.5) (10,0.5)};
\addplot[plotblue] coordinates {(0.1054,0.000) (0.2053,0.005) (0.2057,0.010) (0.2058,0.015) (0.2059,0.020) (0.206,0.025) (0.2061,0.030) (0.2062,0.035) (0.2062,0.040) (0.2063,0.045) (0.2063,0.050) (0.2064,0.055) (0.2064,0.060) (0.2064,0.065) (0.2065,0.070) (0.2065,0.075) (0.2066,0.080) (0.2066,0.085) (0.2067,0.090) (0.2067,0.095) (0.2068,0.100) (0.2069,0.105) (0.207,0.110) (0.207,0.115) (0.2071,0.120) (0.2072,0.125) (0.2072,0.130) (0.2073,0.135) (0.2074,0.140) (0.2075,0.145) (0.2076,0.150) (0.2078,0.155) (0.2079,0.160) (0.2081,0.165) (0.2086,0.170) (0.2148,0.175) (0.25715,0.180) (0.3063,0.185) (0.3066,0.190) (0.3068,0.195) (0.307,0.200) (0.3071,0.205) (0.3073,0.210) (0.3074,0.215) (0.3075,0.220) (0.3076,0.225) (0.3078,0.230) (0.3079,0.235) (0.308,0.240) (0.308,0.245) (0.3081,0.250) (0.3082,0.255) (0.3084,0.260) (0.3085,0.265) (0.30865,0.270) (0.3088,0.275) (0.3088,0.280) (0.3089,0.285) (0.309,0.290) (0.3091,0.295) (0.3092,0.300) (0.3094,0.305) (0.3096,0.310) (0.3098,0.315) (0.3099,0.320) (0.3101,0.325) (0.3104,0.330) (0.3107,0.335) (0.3112,0.340) (0.3118,0.345) (0.3161,0.350) (0.3184,0.355) (0.35915,0.360) (0.4068,0.365) (0.4072,0.370) (0.4075,0.375) (0.4078,0.380) (0.4081,0.385) (0.4082,0.390) (0.4083,0.395) (0.4085,0.400) (0.40875,0.405) (0.4089,0.410) (0.4091,0.415) (0.4093,0.420) (0.4095,0.425) (0.4096,0.430) (0.4098,0.435) (0.41,0.440) (0.4101,0.445) (0.4102,0.450) (0.4104,0.455) (0.4105,0.460) (0.4106,0.465) (0.4108,0.470) (0.411,0.475) (0.4111,0.480) (0.4113,0.485) (0.4114,0.490) (0.4117,0.495) (0.4118,0.500) (0.412,0.505) (0.4122,0.510) (0.4124,0.515) (0.4127,0.520) (0.41295,0.525) (0.4132,0.530) (0.4134,0.535) (0.4137,0.540) (0.4141,0.545) (0.4148,0.550) (0.4154,0.555) (0.4164,0.560) (0.4179,0.565) (0.4194,0.570) (0.4212,0.575) (0.4234,0.580) (0.4345,0.585) (0.46125,0.590) (0.46635,0.595) (0.5083,0.600) (0.5097,0.605) (0.5107,0.610) (0.5116,0.615) (0.5122,0.620) (0.5126,0.625) (0.5129,0.630) (0.5132,0.635) (0.5136,0.640) (0.5138,0.645) (0.5141,0.650) (0.5144,0.655) (0.5149,0.660) (0.5155,0.665) (0.5159,0.670) (0.5161,0.675) (0.5167,0.680) (0.5173,0.685) (0.5177,0.690) (0.5188,0.695) (0.5196,0.700) (0.5207,0.705) (0.5214,0.710) (0.5225,0.715) (0.5252,0.720) (0.5278,0.725) (0.5339,0.730) (0.5543,0.735) (0.5697,0.740) (0.6124,0.745) (0.6141,0.750) (0.615,0.755) (0.6161,0.760) (0.618,0.765) (0.6188,0.770) (0.61975,0.775) (0.6206,0.780) (0.6215,0.785) (0.6222,0.790) (0.624,0.795) (0.6263,0.800) (0.6295,0.805) (0.6373,0.810) (0.6601,0.815) (0.67035,0.820) (0.7101,0.825) (0.7158,0.830) (0.7192,0.835) (0.7212,0.840) (0.7225,0.845) (0.7244,0.850) (0.7261,0.855) (0.7305,0.860) (0.739,0.865) (0.7607,0.870) (0.8172,0.875) (0.8205,0.880) (0.8242,0.885) (0.8261,0.890) (0.833,0.895) (0.8489,0.900) (0.8804,0.905) (0.9182,0.910) (0.9214,0.915) (0.9293,0.920) (0.9356,0.925) (0.9566,0.930) (0.9854,0.935) (1.0208,0.940) (1.0301,0.945) (1.0415,0.950) (1.0877,0.955) (1.1249,0.960) (1.1398,0.965) (1.1771,0.970) (1.2328,0.975) (1.337,0.980) (1.4997,0.985) (1.6295,0.990) (2.0941,0.995) (7.1765,1.000)};
\addlegendentry{IPS, \texttt{verify\_chunk}}
\addplot[plotorange, dashed] coordinates {(0.103,0.000) (0.1035,0.005) (0.1043,0.010) (0.1052,0.015) (0.106,0.020) (0.1066,0.025) (0.107,0.030) (0.1076,0.035) (0.1085,0.040) (0.1091,0.045) (0.11,0.050) (0.1114,0.055) (0.112,0.060) (0.1128,0.065) (0.1136,0.070) (0.1144,0.075) (0.1154,0.080) (0.1163,0.085) (0.1178,0.090) (0.1196,0.095) (0.1227,0.100) (0.1389,0.105) (0.15525,0.110) (0.1698,0.115) (0.19905,0.120) (0.2068,0.125) (0.20795,0.130) (0.2099,0.135) (0.2105,0.140) (0.2116,0.145) (0.2125,0.150) (0.2133,0.155) (0.2142,0.160) (0.2155,0.165) (0.2165,0.170) (0.21745,0.175) (0.2186,0.180) (0.2202,0.185) (0.2213,0.190) (0.222,0.195) (0.2235,0.200) (0.2251,0.205) (0.2258,0.210) (0.2266,0.215) (0.2276,0.220) (0.2288,0.225) (0.2294,0.230) (0.23,0.235) (0.2306,0.240) (0.2312,0.245) (0.2319,0.250) (0.23245,0.255) (0.2335,0.260) (0.2341,0.265) (0.2345,0.270) (0.2353,0.275) (0.2359,0.280) (0.2363,0.285) (0.2371,0.290) (0.23795,0.295) (0.2386,0.300) (0.2393,0.305) (0.2402,0.310) (0.2415,0.315) (0.243,0.320) (0.2441,0.325) (0.2451,0.330) (0.2465,0.335) (0.2478,0.340) (0.2495,0.345) (0.2514,0.350) (0.2536,0.355) (0.2566,0.360) (0.2593,0.365) (0.2624,0.370) (0.2679,0.375) (0.2724,0.380) (0.275,0.385) (0.27915,0.390) (0.2842,0.395) (0.2902,0.400) (0.2951,0.405) (0.299,0.410) (0.3042,0.415) (0.3119,0.420) (0.3161,0.425) (0.3213,0.430) (0.3263,0.435) (0.3299,0.440) (0.3331,0.445) (0.3386,0.450) (0.3439,0.455) (0.3475,0.460) (0.3496,0.465) (0.3523,0.470) (0.3544,0.475) (0.3571,0.480) (0.3607,0.485) (0.3627,0.490) (0.3667,0.495) (0.3708,0.500) (0.3727,0.505) (0.3752,0.510) (0.3776,0.515) (0.37925,0.520) (0.3812,0.525) (0.3834,0.530) (0.3858,0.535) (0.3879,0.540) (0.3896,0.545) (0.3913,0.550) (0.3925,0.555) (0.3939,0.560) (0.3955,0.565) (0.39785,0.570) (0.3997,0.575) (0.4009,0.580) (0.4025,0.585) (0.4045,0.590) (0.4062,0.595) (0.4088,0.600) (0.4115,0.605) (0.4143,0.610) (0.417,0.615) (0.4215,0.620) (0.4238,0.625) (0.4262,0.630) (0.4286,0.635) (0.4327,0.640) (0.435,0.645) (0.4387,0.650) (0.4429,0.655) (0.4476,0.660) (0.45185,0.665) (0.45815,0.670) (0.4627,0.675) (0.4667,0.680) (0.4705,0.685) (0.474,0.690) (0.4769,0.695) (0.4821,0.700) (0.48605,0.705) (0.4907,0.710) (0.4957,0.715) (0.5006,0.720) (0.5046,0.725) (0.5099,0.730) (0.5148,0.735) (0.52,0.740) (0.5263,0.745) (0.5313,0.750) (0.5406,0.755) (0.5439,0.760) (0.5546,0.765) (0.56245,0.770) (0.5682,0.775) (0.5762,0.780) (0.5871,0.785) (0.5991,0.790) (0.6067,0.795) (0.6237,0.800) (0.6363,0.805) (0.6458,0.810) (0.6598,0.815) (0.6746,0.820) (0.6913,0.825) (0.702,0.830) (0.7152,0.835) (0.7347,0.840) (0.7555,0.845) (0.7781,0.850) (0.795,0.855) (0.8089,0.860) (0.83935,0.865) (0.8675,0.870) (0.8863,0.875) (0.9058,0.880) (0.93145,0.885) (0.9564,0.890) (0.9752,0.895) (0.9912,0.900) (1.0088,0.905) (1.0428,0.910) (1.0765,0.915) (1.1,0.920) (1.1315,0.925) (1.1601,0.930) (1.1875,0.935) (1.2264,0.940) (1.2689,0.945) (1.3144,0.950) (1.3811,0.955) (1.4424,0.960) (1.489,0.965) (1.5743,0.970) (1.6523,0.975) (1.7535,0.980) (2.0087,0.985) (2.2181,0.990) (2.7806,0.995) (8.0974,1.000)};
\addlegendentry{PIDE MCP, edit and settle}
\end{axis}
\end{tikzpicture}
\caption{Empirical cumulative distribution
  $F(t) = \Pr[\,T \le t\,] = \frac{1}{n}\,|\{\,i : T_i \le t\,\}|$ of the
  latency $T$ of one call at one worker, over the same $n = 2\,966$ chunks
  under both tools, $T_i$ being the latency of chunk $i$ under the tool, the
  median over the repeats that ran it; each curve passes through every
  0.5\,\% quantile (Appendix~\ref{app:bench-training}). The vertical runs
  are the 0.1\,s step of both tools, the interval at which IPS re-reads
  the node and at which our harness polls PIDE MCP; the dotted line is
  $F(t) = 0.5$, the
  median, at 0.41\,s under IPS and 0.37\,s under PIDE MCP.}
\label{fig:latency}
\end{figure}

\paragraph{Latency.} Cold starts do not separate the two tools. A start
takes 2 to 7\,s depending on the heap, and the gap between the tools'
medians, 0.2 to 0.8\,s, is smaller than the spread of IPS's own start
times, about 1.5 to 2\,s from the median to the 95th percentile
(Table~\ref{tab:capacity}). For the latency of one call we paired the
chunks of the one-worker runs above: 2\,966 chunks that both tools
proved, each timed from the call to the moment the client knew it had
settled, at the same 0.1\,s resolution
(Appendix~\ref{app:bench-training}); Figure~\ref{fig:latency} shows the
two distributions. Checked one after another, they took IPS 1\,479\,s
and PIDE MCP 1\,443\,s. The difference is 2.4\,\%, or 12\,ms per call,
which is small against the 0.1\,s polling interval. The
median call is 0.41\,s under IPS and 0.37\,s under PIDE MCP, within one
polling interval, as are 57\,\% of the pairs; a single command, the floor of a call,
costs 0.34\,s under IPS and 0.31\,s under PIDE MCP on average, and chunks
of ten or more commands cost the same under both (0.66\,s). The IPS call
does more for the time, as the paragraph above says, and an event-driven
client would shorten both.

\paragraph{Under a memory limit.} We make no comparison under a memory
limit. PIDE MCP at that commit leaves admission to the deployment, as
the previous subsection says, so a run under a limit tests the kernel
and not the tool. The run we made under topology~B is in the results
repository.

\paragraph{Isolation.} With one prover among 8 killed, the other 7
sessions answered under all three arrangements. PIDE MCP reports the
killed session as no longer running, where IPS names the cause
(Section~\ref{sec:serving-measured}). Under topology B all eight MCP
processes survived
their provers. On this test the tools are equal.

\subsection{Capabilities}

\paragraph{PIDE MCP at the measured commit, not IPS.} A project of many
theories and auxiliary files with dependency resolution; an edit tool
that writes files, with messages at line and column; a visible range
around what the agent reads or edits, on which PIDE runs its print
functions; cancellation of a long-running tool call and progress
notifications; entity search over a range; an extension mechanism for
tools through Isabelle services; and three shipped agent skills. The
first is a matter of what a session is; the others are tools.

\paragraph{IPS, not PIDE MCP at the measured commit.} A pool with
leases, reuse, eviction and a memory admission gate;
rollback of a failed chunk as a unit, and checkpoints; a
verification call with explicit semantics (success, open
proof, \code{sorry}) and a batch build as arbiter; a wall-clock
budget on every call; batch fan-out across the pool; an HTTP API beside
MCP; input hardening (an ML and file-IO denylist, path roots);
Prometheus metrics and probes; and a test suite. The pool with its
leases and its gate is the resource-management layer. The other items
are tools that need neither leases nor appended text, and they could be
written for another PIDE frontend.

\paragraph{Both.} Direct PIDE frontends on headless sessions with chosen
logic and imports; per-command status; state queries that do not edit
the document; Sledgehammer; MCP on standard input and output.

\section{Limitations and roadmap}
\label{sec:limitations}

\subsection{Limitations}

The limits are of three kinds: of what a session is, of the deployment,
and of the measurements. Two further sets are reported elsewhere.
Section~\ref{sec:comparison} lists the tools PIDE MCP had and IPS lacks,
cancellation, progress and an extension mechanism among them. And two
weaknesses that the measurements found are reported where they were
measured: session starts slow down as workers are added
(Section~\ref{sec:training-measured}), and a long rewriting loop leaves
memory in the prover (Section~\ref{sec:checks}); the pool does not
recycle such a session, so the client must close it.

\paragraph{What a session is.} A session's imports are fixed when it is
created, and it holds one theory with no file behind it. So changing
imports means a new session, and project imports must be session-qualified
theory names; there is no multi-file project scope; diagnostics are located at
a command, not at a line within it; and the connector edits no file: the
file MCP server observes a file the agent edits with its own tools, and
the stepwise server has no file at all. This is the price of the
decision that lets sessions be pooled (Section~\ref{sec:comparison}).

\paragraph{Deployment.} There is no authentication. The trust boundary
is the network: any client that can reach the server can acquire
sessions, and the lease check keeps clients apart but does not identify
them. IPS is meant to run on a trusted network or behind an
authenticating proxy (Appendix~\ref{app:deploy}). One gateway JVM holds
every session's backend and grows with the pool (3.6 to 5.1\,GB over 32
sessions); a crash of it loses every session at once.

\paragraph{The measurements.} Four things bound what the numbers show.
\begin{enumerate}
  \item \textbf{They are tied to a build.} Every figure in
    Sections~\ref{sec:training} to~\ref{sec:comparison} is for the
    Isabelle build, host and commits in Appendix~\ref{app:bench}. They
    were taken on the second release candidate of Isabelle2026 and were
    not re-taken on the final release.
  \item \textbf{PIDE MCP was measured at one commit.} No number was
    re-taken on PIDE AI, and three changes made since that commit bear
    on the results (Section~\ref{sec:pide-version}).
  \item \textbf{The replay baseline is IPS itself.} The branch-cost
    comparison replays a prefix in a new node or a new session of IPS,
    the cost of a branch from a state nobody saved. A REPL on the same
    host would make the comparison direct, and one that keeps states
    (Section~\ref{sec:training-req}) would put a saved-state branch
    beside the restore. We ran none: PISA dates from 2021, qIsabelle
    stopped at Isabelle2024, and I/R, which is built for agents and not
    for training loops, we did not try. The PIDE MCP figures beside IPS's
    in Section~\ref{sec:comparison} are throughput at $W$\,=\,1 and 8,
    not branch cost, so the branch-cost comparison has no external
    baseline.
  \item \textbf{The perspective cost was reproduced on IPS, not isolated
    in PIDE MCP.} Making a theory fully visible adds 0.2--0.4\,GB to an
    IPS prover on HOL, which matches PIDE MCP A's gap there, but PIDE
    MCP itself was not re-run with a narrower range, and HOL-Library and
    HOL-Analysis were not checked. Our harness writes the whole theory
    in one edit, so narrowing \path{pide_mcp_range_context} alone would
    not change what is visible; a harness that edits in pieces would. A
    re-run on PIDE AI, where visibility is the agent's choice
    (Section~\ref{sec:pide-version}), would test the attribution
    directly; we have not done it.
\end{enumerate}

\subsection{Roadmap}

A strict verdict is the first item, and authentication the first for
the multi-agent serving scenario. All but the last leave IPS's session
layer as it is.
\begin{itemize}
  \item \textbf{A strict verdict.} Reject any theorem whose oracle set is
    not empty, which Isabelle's \code{thm\_oracles} reports and which
    covers \code{sorry} and every other oracle, and check the statement
    against a reference, as AXLE's strict verifier does for Lean.
  \item \textbf{Authentication with per-client quotas} on sessions and
    budgets, so that a shared host can tell its clients apart and bound
    each one.
  \item \textbf{Agents over the AST.} AoA~\cite{xu2026aoa} edits a
    Minilang proof tree by node rather than Isar text and runs today on
    Isa-Mini's own REPL server; a Minilang document mode in IPS, with
    node-level edits and per-node verdicts, would give such agents the
    pool, the budgets and the checkpoints.
  \item \textbf{A command-line client.} An agent that has a shell can
    call a command where it would call an MCP tool. PIDE AI ships
    command-line clients for its command protocol, and AutoCorrode's IC2
    drives headless sessions with
    \code{isabelle ic2}~\cite{awslabs2026autocorrode}. The HTTP API
    already carries every IPS operation, so the client is a thin
    wrapper; what needs design is where the lease is kept between
    invocations.
  \item \textbf{Consolidation with PIDE AI.} IPS and PIDE AI each wrote
    a Scala layer on Isabelle's session class, and below the session
    model the two now do the same things: a node checked without a
    visible range, proof states read from the prover's state messages,
    Sledgehammer as an overlay query. Sharing can start there, with
    IPS's query operations as a component either system loads and the
    harness of Appendix~\ref{app:bench} as a common benchmark. The next
    step is a trial: one IPS operation, a step under a budget that is
    rolled back when the budget runs out, written as a PIDE AI tool
    extension and measured. If the trial goes well, the pool, the leases
    and the memory gate could run on PIDE AI sessions in place of IPS's
    own backend. The two session models differ (where a node's text
    comes from, how a session gets its imports, what a timeout does), so
    we think a merge possible and have not costed it.
\end{itemize}

\section{Conclusion}
\label{sec:conclusion}

IPS is a direct PIDE frontend with a resource-management layer: a pool
of warm sessions, each leased to one client at a time, admitted
by memory, with a budget on every call. In architecture it does not
differ from PIDE MCP, and what a training loop gets from it, rollback,
exact restore, parallel checking within a chunk and a batch build as
arbiter, are tools that could be written for another PIDE frontend.
What is particular to IPS is narrower. Sessions with the same imports
are interchangeable and can be pooled. In exchange a session cannot
hold a project of several theories, and diagnostics are located at a
command and not at a line. And the layer above the sessions is built
and measured: a known cost per session, a gate that refuses with a
reason, and isolation by lease. On the questions we measured, memory
per session, throughput, conduct under a limit and isolation, IPS does
what it was built for. PIDE MCP at commit \code{dd46315}, holding its
sessions in one process, is within 0.4\,GB of it per session and level
with it in throughput at one worker. The three-fold gap in memory per
session, which appears when each agent runs its own PIDE MCP process,
is the price of a JVM per agent, and PIDE AI no longer requires one.
On the questions we did not measure, those of the interactive serving
scenario, PIDE MCP is the more capable tool, and nothing in our
measurements says otherwise. A group running both would use PIDE AI
for interactive work and IPS in the training cluster; the two share
what lies below the session model, so a consolidation is possible
(Section~\ref{sec:limitations}). The measurements are repeatable from
Appendix~\ref{app:bench} and the results repository.

\section*{Acknowledgements}
We thank Kevin Kappelmann for fruitful discussions and comments on an early draft of this report, and Lawrence Paulson for pointing us to the agentic workflow of AutoCorrode. The Isabelle Pool Server was developed as part of the Sketchpad project, supported by Renaissance Philanthropy’s AI for Math Fund.

\bibliographystyle{plainnat}
\bibliography{references}

\appendix

\section{IPS in PIDE's terms}
\label{sec:pide-terms}

This appendix describes the service of Section~\ref{sec:arch} for a reader who knows PIDE~\cite{wenzel2014pide}. IPS is a direct PIDE frontend. A session is a \code{Headless.Session}, the class behind \code{isabelle server}; the backend starts it, loads its imports and edits its document through the Isabelle/Scala API. The session's theory is one node of that document. Its text is held in memory, with no file behind it, and the server appends to it or replaces it. A \emph{chunk} is one insert at the end of the node, and a rollback is the edit that removes it. The node is \emph{required} with an empty visible range, its \emph{perspective}, so PIDE checks every command, reports errors and warnings, and runs no print function. Proof states, facts in scope and Sledgehammer are query operations run as overlays on the node's last command and removed afterwards; the document is never edited for them. A call returns when the submission has \emph{settled}. A \emph{checkpoint} records the node's text and the status of its commands; a \emph{restore} is the edit that takes the text back, so PIDE's incremental checking keeps everything before the edit point and re-checks nothing. A restore call returns when its edit is accepted, before the document has settled; the next call waits until it has, which is why Section~\ref{sec:training-measured} counts that settling delay in the branch cost. Proof checking is parallel inside a chunk: in a headless session any \code{parallel\_proofs} above 0 runs at Isabelle's interactive level, which forks the proofs of a command, so the value the experiments set (2) is the same checking as 1.

\section{Quickstart and API summary}
\label{app:quickstart}

\paragraph{Running the server.} IPS ships as one container image
parameterised by Isabelle version, with a setup script for Docker and one
for a native Linux host (optionally as a systemd unit). On a machine with
Docker:
\begin{verbatim}
git clone https://github.com/Deep-LCF/Isabelle-Pool-Server.git
cd Isabelle-Pool-Server
./deploy/setup.sh --verify     # writes .env, builds, starts, proves True
curl http://localhost:8000/readyz   # 200 once the gateway JVM is up
./deploy/native_setup.sh --system --systemd --verify  # the same, no Docker
\end{verbatim}
The first build downloads Isabelle and compiles the backend (10--30\,min);
the last line is the native install. The configuration is a set of
environment variables in \code{.env} (Appendix~\ref{app:deploy}).

\paragraph{A session from Python.} The client is a separate package
(\texttt{pip install -e ./client}, \code{httpx} only). Every call on a session
carries the lease returned when the session was acquired.
\begin{verbatim}
from client.async_client import PoolAsyncClient

async with PoolAsyncClient("http://localhost:8000") as client:
    s = await client.acquire_session(theories=["Main"], field="HOL")
    sid, lease = s["session_id"], s["lease_id"]
    try:
        await client.enter_theory(sid, "T", imports=["Main"], lease_id=lease)
        cp = await client.save_checkpoint(sid, lease_id=lease)
        r = await client.verify_chunk(
            sid, 'lemma l: "x + 0 = x" by simp', timeout=10.0, lease_id=lease)
        proved = r["success"] and not r["proof_open"] and not r["used_sorry"]
        if not proved:
            await client.restore_checkpoint(
                sid, cp["checkpoint_id"], lease_id=lease)
    finally:
        await client.release_session(sid, lease_id=lease)
\end{verbatim}
\code{verify\_chunk} returns the verdict flags and the per-command
report:
\begin{verbatim}
{"success": true, "proof_open": false, "pending_qed": false,
 "used_sorry": false, "timed_out": false, "stuck_line": null,
 "execution_time": 0.41,
 "commands": [{"index": 0, "line": 1, "kind": "lemma", "status": "ok",
               "messages": []}, ...]}
\end{verbatim}
On a budget expiry \code{timed\_out} is true and \code{stuck\_line}
names the command still running; \code{pending\_qed} says the goals are
discharged and only the closing \code{qed} is missing.
\code{load\_document} replaces the whole document, \code{rollback}
removes the last submission, and \path{verify_bigstep_text} is the
batch build.

\paragraph{The stepwise MCP server (13 tools).} \code{enter\_theory},
\code{verify\_chunk}, \code{proof\_state}, \code{source},
\code{diagnostic}, \code{sledgehammer}, \code{checkpoint}, \code{restore},
\code{rollback}, \code{close\_theory}, \code{verify\_batch},
\code{list\_my\_sessions}, \code{memory\_status}.
\code{verify\_chunk} is the only tool that executes Isar; it returns the
flags above as one text line, then one line per failed or running
command (every command with \texttt{detail=true}), and a note when the
proof is still open or only \code{qed} is missing.

\paragraph{The file MCP server.} The agent edits a theory file
with its own tools; the server re-reads it on every call and keeps the
session in step. Tools cover open, close and sync; diagnostics, goal and
proof state at a line; source, hover and definition; local and global
facts; query and Sledgehammer; checkpoint, restore, rollback and history;
multi-attempt and run-code on scratch sessions; and heap build and
status. The naming follows
lean-lsp-mcp~\cite{dressler2026leanlspmcp}.

\paragraph{The HTTP API.} Served with an OpenAPI document at
\code{/openapi.json}. The routes a client uses in order: acquire or
create a session; submit a command, a chunk or a whole document; read
state, goals or facts; checkpoint, restore, rollback; release. The
whole-file check is a stateless route that takes a theory and returns a
verdict; batch fan-out is the stepwise MCP's \code{verify\_batch}, not a
route. \code{/api/v1/memory} reports the memory
facts behind every admission decision.

\paragraph{Repeating the measurements.} Appendix~\ref{app:bench} says
where the scripts and the raw results are; \code{SERVER\_GUIDE.md} in
\path{evaluation/training_metrics/} is the step-by-step procedure for a
fresh Ubuntu host.

\section{Experimental details}
\label{app:bench}

Everything here is reproducible from the IPS repository
(\path{evaluation/memory_scaling/} for the serving runs,
\path{evaluation/training_metrics/} for the training runs; each one
\code{config.yaml}, numbered scripts and a deterministic analysis). The
raw data is in the results
repository\footnote{\url{https://github.com/FullBackward/Isabelle-Pool-Server-results}}:
every cell's manifest (phases, per-session timings, environment,
workload checksum, the tool's own report), its per-second samples, the
per-cell tables and the plots; the memory-scaling tree also as release
asset \path{memscale-2026-10-21b} of the main repository. This appendix
keeps the method and the two series the body's figures and fits are
drawn from. The per-cell tables (E1 medians per cell, E3 per session,
the E2 and E4 records) are \path{results/report-raw-tables.md} in the
results repository.

\paragraph{Environment.} One x86-64 host with 32 logical CPUs and
128\,GB of memory, Ubuntu with cgroup v2, nothing else running; both
tools native, as one unprivileged user, from one Isabelle2026
installation (the second release candidate, Section~\ref{sec:limitations})
with the same 9\,GB memory cap on every prover process
(\texttt{ML\_OPTIONS --maxheap 9216}).
IPS at commit \code{6e4dd6f} for the memory-scaling runs and
\code{1fe270f} for the training runs (the census at \code{28600bb} for
82 theories and \code{cb03c0e} for the 11 whose blocks a splitter fix
re-cut, a server-side change judged not to touch a replay; the analysis
at \code{e144f97}), memory management on at
its defaults, pool size 40, no pre-warmed sessions,
\code{parallel\_proofs=2}; the training runs with a 1.5\,GB estimate per
HOL-Analysis session and four batch builds at a time; a fresh server in
its own systemd scope for every memory-scaling cell. PIDE MCP at commit
\code{dd46315} of 19 October 2026 (server 0.1.0; the repository is now
\path{kappelmann/isabelle-pide-ai}), reached over MCP on standard input
and output, with its default options, among them
\path{pide_mcp_range_context}\,=\,50, and \code{parallel\_proofs=2} on
its sessions in the training rows.

\paragraph{Memory-scaling workloads.} One fixed theory per heap, loaded
whole into every session: \code{CTL.thy} (HOL, 288 commands),
\code{Termination.thy} (HOL-Library), \code{Sum\_Topology.thy}
(HOL-Analysis, 163 lines, importing
\path{HOL-Analysis.Abstract_Topology}). The re-check inserts
\texttt{lemma memscale\_p5\_done: "True" by simp} after the first
paragraph under the same theory name, so every later command is
re-checked on the incremental path.

\paragraph{Memory measure.} We report the sum of PSS over the tool's
processes, read once a second from \path{/proc/<pid>/smaps_rollup}: PSS
counts a page shared by $n$ processes as $1/n$ for each, so the sum is
the physical memory the tool occupies whatever is shared. Sharing
between provers is small, since Poly/ML reads the heap image into private
memory. Host used
(\code{MemTotal} $-$ \code{MemAvailable}) is the independent check: its
per-session slope agrees with the PSS slope within 2\,\% in every series
of Table~\ref{tab:capacity}, and its intercept sits 0.54--0.67\,GB below
the PSS intercept in every series, most likely the file-backed pages the
JVMs keep resident, which PSS counts and \code{MemAvailable} treats as
reclaimable; the table therefore reports the PSS fit, and the host fits
are in the results repository. The cgroup counters of the systemd scope the
tool runs in (\code{memory.current}, \code{memory.peak},
\code{oom\_kill}) are what a memory limit is enforced on and carry the
E3 result.

\paragraph{E1, capacity.} For each tool, heap and $k$ in
$\{1,2,4,8,16,32\}$, three repeats: P0 baseline (60\,s; 10\,s after the
first cell of a series), P1 start $k$ sessions, P2 idle 60\,s, P3 load the
theory in each session, P4 settle 60\,s, P5 re-check all $k$ concurrently,
P6 stop and a 30\,s tail. Steady state is the median of the last 30\,s
of a phase; host used is the delta over the cell's own P0; the fits of
Table~\ref{tab:capacity} are least squares over the 18 cells of a
series. The lower row of Figure~\ref{fig:capacity} is the P5 series, the re-check peak;
the per-cell values are \path{results/tables/cells.csv} in the results
repository.

\paragraph{Perspective check.} To test the attribution of
Section~\ref{sec:comparison}, IPS was rebuilt with its nodes' visible
range set to the whole text (the two places the backend sets the
perspective) and run in its container on a 16\,GB machine with
Isabelle2026. Four HOL sessions each loaded the E1 HOL theory through
\code{load\_document} and settled for 60\,s; then each prover's PSS was
read from \path{/proc}. The theory's one \code{axiomatization}, a constant
declaration without axioms, was written as \code{consts}, because the
command guard now refuses the word. Empty range: 816--819\,MB in all four provers. Full
range, two repeats: 1\,007--1\,176 and 665--1\,251\,MB. Proof-state
output (\path{editor_output_state}) was off throughout, as in both
tools' defaults. The script and the three samples are
\path{results/perspective-check/} in the results repository.

\paragraph{E2, E3, E4.} E2: 4 IPS sessions, released and re-acquired
with the same imports for 3 rounds. E3: IPS sessions added one at a time,
the server in a systemd scope with \code{MemoryMax=48G}, until a refusal,
a failed load or an \code{oom\_kill}; then half
the sessions released and one more added. E4: 8 HOL sessions, the newest session's
prover killed with SIGKILL, every session asked to answer within 30\,s
(IPS: a re-check through \code{load\_document}; PIDE MCP:
\code{get\_state}). Section~\ref{sec:serving-measured} reports the
outcomes.

\paragraph{Training workloads.}
\label{app:bench-training}
The raw results of the runs of Section~\ref{sec:training-measured},
every cell's manifest, per-block records, memory and CPU samples and the
server's reports, are the \code{training-metrics/} tree of the results
repository.

\emph{Corpus and census.} The 94 theories of HOL-Analysis in
Isabelle2026 less the collector \code{Analysis.thy}, 93, with their
imports qualified, each split into top-level
blocks (11\,984; a block is one top-level command with its proof). The
census replays every theory once at one worker, each on a session with the
theory's imports, with a 600\,s budget per block; it proved all 11\,984
blocks, and its per-block times and command counts key every later
measurement. A block's budget in the experiments is $\min(600,
\max(30, 2n))$\,s for $n$ census commands.

\emph{T1 throughput, T2 latency.} $W$ workers (1 to 32, three repeats),
8 theories each from a seeded assignment that is a prefix across $W$, each
theory on a fresh session acquired with its imports, replayed block by
block with \code{verify\_chunk}; the clock runs from the first theory to
the last. Table~\ref{tab:t1-series} is the full series. T2, the latency of one call, is read from the T1 cells. The PIDE MCP runs
use the same assignment at $W$\,=\,1 and 8 through a loop of our own: a
session per theory, a file from \code{create\_scratch}, one \code{edit}
per block carrying the whole text, and \code{get\_state} every 0.1\,s
until the block has settled. A block that does not settle has to be edited out by the client, which
our driver did not do, so such a block ended its theory (one per cell).
Isabelle requires a theory to bear the name of its file, so each theory
took the scratch file's name, and the 1\,\% of blocks that refer to
their own theory by its library name fail for that reason; the choice
of file was the driver's, since PIDE MCP does not name theories. The
loop also closes the text with \code{end} after every edit, so a block
that opens an \code{instantiation} is counted as failed, and so are
some blocks after it in the same theory; the two cases together are
under 1\,\% of the blocks. For the
latency comparison of Section~\ref{sec:comparison} its blocks pair with
IPS's by block hash over the one-worker cells, a block's time the median
of the repeats that ran it, and only blocks proved under both tools count:
2\,966 of 2\,985, one having timed out and 18 being proved under one tool only.
\emph{T3 branch cost.} 200 proof states, a third each early, in the middle
and late in their theories, each on a session created for it: checkpoint,
verify the next block, three restores each followed by the block, two
rollbacks each followed by the block, then the prefix replayed in a new
theory node and in a new session. \emph{T4 verdict flags.} Part A: each theory
as written, the census verdict against \path{verify_bigstep_text}
(\code{isabelle build} on the heap). Part B: 200 census-proved theorem
blocks, seeded, in four forms (unchanged, \code{sorry}, truncated before
the closing step, first fact name undefined), checked in-session at the
block's position under its budget and rolled back, and in a batch build with the form
spliced into its theory, one mutation per build. \emph{T5 bounded step.}
50 looping steps of each kind, a rewriting loop (\texttt{apply (subst add.commute)+} on an
additive goal) and a \code{metis} call with up to 56 facts on a false goal, placed before
census-proved theorems of 9 theories,
tried at 5\,s; of those that looped, up to 10 of each kind tried again at 30\,s
and up to 3 at 120\,s; a 1.5\,s pause after every attempt for the memory
sampler; at $W$\,=\,8, two looping steps per worker, under 120\,s, among the worker's
theories (16, of which one rewriting step did not loop). \emph{T6 rollout loops.} $W$ workers (1, 8, 32; two repeats), three
theories each from the T1 assignment; whole-proof: the first 8 proved
theorems of each theory, 8 candidates each (the corpus proof, \code{by
auto} and \code{by simp} twice, \code{sorry}, the proof truncated, a method
that fails at once) under 20\,s, one that stays in the document discarded by a restore;
step-level: the first 4 proved theorems of 2 to 40 proof commands, at each
command two alternatives from \{\code{apply auto}, \code{apply simp},
\code{apply -}, \code{by auto}, \code{by simp}\} under 10\,s before the
corpus command. Every cell polls the server's memory tier and samples
memory and CPU per process once a second.

\begin{table}[!htb]
\centering\footnotesize
\setlength{\tabcolsep}{2pt}
\begin{tabular}{@{}r rr rr rr r rr r@{}}
\toprule
W & \shortstack{chunks/min\\verify only} & \shortstack{repeats\\min--max} & \shortstack{commands\\per min} & \shortstack{chunks/min\\with session starts} & \shortstack{acquisition\\share} & \shortstack{census\\work rate} & \shortstack{slowdown\\vs census} & \shortstack{p50\\s} & \shortstack{p95\\s} & \shortstack{PSS peak\\GB} \\
\midrule
1  & 132   & 108--132 & 1\,607  & 113 & 0.10 & 0.97 & 1.02 & 0.41 & 0.92 & 6.5 \\
2  & 281   & 230--286 & 3\,713  & 183 & 0.12 & 1.44 & 1.04 & 0.41 & 0.92 & 7.5 \\
4  & 474   & 469--503 & 6\,576  & 297 & 0.12 & 2.58 & 1.08 & 0.41 & 1.06 & 11.3 \\
8  & 927   & 902--975 & 12\,501 & 497 & 0.14 & 4.45 & 1.17 & 0.42 & 1.22 & 16.0 \\
16 & 1\,677 & 1\,558--1\,718 & 22\,353 & 789 & 0.30 & 7.07 & 1.62 & 0.47 & 1.41 & 26.5 \\
32 & 3\,181 & 3\,162--3\,214 & 44\,877 & 899 & 0.64 & 8.33 & 3.13 & 0.52 & 1.50 & 28.3 \\
\bottomrule
\end{tabular}
\caption{T1 at every $W$: the median of three repeats, with the repeats'
  range of the verify-only rate (no call timed out, no theory skipped).
  Census work rate: the census time of the cell's theories over the
  cell's wall time, the number of single-worker replays the cell is
  worth; speedup and efficiency in the body are on it. Slowdown: the
  cell's replay time over the census time of the same theories,
  acquisition included.}
\label{tab:t1-series}
\end{table}

\begin{table}[!htb]
\centering\small
\begin{tabularx}{\textwidth}{@{}>{\raggedright\arraybackslash}X l r >{\raggedright\arraybackslash}X l@{}}
\toprule
Proof as submitted & flag it targets & n & in-session & batch build \\
\midrule
Proof replaced by \code{sorry} & \code{used\_sorry} & 200 & \code{used\_sorry} fired in 200 & rejected 200 \\
Last closing step removed & \code{proof\_open} & 200 & \code{proof\_open} fired in 200 & rejected 200 \\
A fact name undefined & \code{success} & 197 & \code{success} false in 197, 13 of them at the budget & rejected 197 \\
Block unchanged & none & 200 & no flag fired in 200 & accepted 200 \\
Theory as written & none & 93 & no flag fired in 93 & accepted 93 \\
\bottomrule
\end{tabularx}
\caption{Each verdict flag against the failure it is meant to catch, with
  a batch build of the same text as the reference. A flag that
  fires rejects the proof in-session. The mutations change the proof text
  only; statements and axioms are as in the library. Three blocks have no
  fact name to replace.}
\label{tab:agreement}
\end{table}

\begin{table}[!htb]
\centering\small
\begin{tabular}{@{}l r r r rr r@{}}
\toprule
Looping step & Budget & Attempts & Returned at budget & \multicolumn{2}{c}{Rollback p50 / p95} & Memory left, p50 \\
\midrule
\code{metis}, up to 56 facts & 5\,s & 50 & 50 & 0.07\,s & 0.45\,s & 4\,MB \\
 & 30\,s & 10 & 10 & 0.30\,s & 0.60\,s & 127\,MB \\
 & 120\,s & 3 & 3 & 0.28\,s & 0.35\,s & 1\,MB \\
\code{subst} rewrite loop & 5\,s & 38 & 38 & 0.05\,s & 0.52\,s & 4\,MB \\
 & 30\,s & 8 & 8 & 0.54\,s & 4.21\,s & 1\,181\,MB \\
 & 120\,s & 1 & 1 & 0.35\,s & 0.35\,s & 7\,116\,MB \\
\bottomrule
\end{tabular}
\caption{Looping steps under a budget, one worker. Rollback is the wall
  time past the budget until the call returned with the step rolled back.
  Memory left is the prover's resident set after the attempt minus before
  it. Twelve of the 50 rewriting steps did not loop and are left out.}
\label{tab:bounded}
\end{table}

\section{Deployment notes and security envelope}
\label{app:deploy}

\paragraph{Sizing.} About 30\,GB of disk for Isabelle and the heaps;
memory as measured in Section~\ref{sec:serving-measured}: the pool's fixed
cost plus one per-session cost of Table~\ref{tab:capacity} for each
concurrent session, with the admission gate's margin on top. The pool
size, the number of
sessions started at boot, the idle and lease timeouts, the memory
thresholds and session estimates, the prover memory cap and the per-call
budgets are environment variables documented in the repository's
\code{.env.example}.

\paragraph{Setup.} The Dockerfile selects the native architecture and
falls back across mirrors; Appendix~\ref{app:quickstart} gives the setup
scripts.
Readiness and liveness are exposed as HTTP probes; metrics are exposed for
Prometheus, with a Grafana dashboard in the repository. The memory gate
reads the server's own cgroup, so it sees a Docker memory limit and a
systemd \code{MemoryMax} alike.

\paragraph{Security envelope.} There is no authentication
(Section~\ref{sec:limitations}), and the administrative page inlines its
token, a known and deferred issue. Every endpoint that accepts Isar text rejects ML, setup, file-IO and
\code{axiomatization} commands unless the operator switches the policy
off. The
diagnostic endpoint accepts only an allowlist of keywords. Heap project
paths are confined to configured roots, and destructive heap operations
require an administrative token. Measured against the MCP specification's
requirements for input validation, access control and rate
limiting~\cite{mcp2025spec}, IPS meets the first in part and neither of
the others. PIDE MCP states that it includes no safety measures and refers
to a separate set of sandboxed setups~\cite{schaeffeler2026safe}. PIDE
AI's TCP port, like the IPS port, has no authentication.

\paragraph{Where to look.} The backend (session start, document edits,
overlay queries, the per-command report) is \path{server/repl}; the
HTTP layer, the pool and the memory monitor \path{server/app}; the two
MCP servers \path{mcp_servers}; the experiments \path{evaluation}.
\path{docs/DESIGN_CHOICES.md} in the repository explains how the server
keeps the guarantees of Section~\ref{sec:arch}: the tiers, the owned
document, verification, the overlay queries, and the pool with its
admission gate.

\section{A worked example: agents in the loop (anecdote)}
\label{app:example}

The Isabelle-Humanize
project\footnote{\url{https://github.com/FullBackward/Isabelle-Humanize}}
runs a builder/reviewer loop (the RLCR flow of
humanize2\footnote{\url{https://github.com/humanfia/humanize2}}) on
competition problems. Two agent seats (opencode on DeepSeek V4 Pro in
both roles) talk to IPS through the file MCP server over streamable HTTP,
and the flow uses the IPS whole-file check as the arbiter: a problem
counts as solved only when \code{isabelle build} passes with no
\code{sorry} or \code{oops} and the target theorem present. Three
problems were solved end to end: Putnam 2023 A1 (2 review rounds, 3.3\,h
wall), Putnam 2023 B5 (5 rounds, 14.3\,h, with
one regression and a replan) and IMO 2026 P1 (one round, a
1059-line theory with all five \code{sorry}s discharged, 4.2\,h). A rerun of IMO 2026 P1 on the IPS build with incremental document
sync and the memory work of this report halved the wall time (7\,391
against 14\,949\,s) and the agent-active time (117 against 248\,min),
cut MCP timeouts from 12 to 0 and diagnostics calls from 47 to 18 at a
median of 0.3\,s, and the arbiter build from 44 to 11\,s. The project's
own caveats apply: one run per problem, a prompt rule that changed between
the two IMO runs, and a
fourth ``solve'' (Putnam 2023 B6) excluded because the benchmark's Isabelle
statement is degenerate. This is an example, not an evaluation: a few
problems with one model measure the agent and the prompt as much as the
connector. It shows that a service built for the first two scenarios
also serves the third, and that the service's latency reaches the
agent's wall
clock.

\end{document}